\documentclass[11pt,english]{article}
\usepackage{graphicx}
\usepackage{array} 
\usepackage{bbm}
\usepackage{booktabs}
\usepackage{caption}
\usepackage{color}
\usepackage{enumerate}
\usepackage{comment}
\usepackage{hyperref}
\usepackage{multicol}
\usepackage{multirow}
\usepackage{natbib}
\usepackage{setspace}
\usepackage{titlesec}
\titleformat{\section}
{\normalfont\fontsize{12}{15}\bfseries}{\thesection}{1em}{}
\usepackage{url}
\usepackage{amsmath,amsfonts,amsthm,amssymb}
\usepackage{subcaption} 
\usepackage{xfrac}
\usepackage{makecell}
\usepackage{geometry}
\usepackage{siunitx}
\usepackage{enumitem}
\usepackage{booktabs}
\usepackage{float}
\DeclareMathOperator{\EX}{\mathbb{E}}

\newcommand{\tildea}[1]{\stackrel{\sim}{\smash{#1}\rule{0pt}{1.1ex}}}
\newcommand{\hata}[1]{\stackrel{\wedge}{\smash{#1}\rule{0pt}{1.1ex}}}

\title{Inverse and Quanto Inverse Options in a Black-Scholes World}
\author{\large Carol Alexander,\footnote{University of Sussex Business School. Email: \href{mailto:c.alexander@sussex.ac.uk}{c.alexander@sussex.ac.uk}}  \,    Ding Chen\footnote{University of Sussex Business School. Email: \href{mailto:ding.chen@sussex.ac.uk}{ding.chen@sussex.ac.uk}} \, and  Arben Imeraj\footnote{University of Sussex Business School. Email: \href{mailto:a.imeraj@sussex.ac.uk}{a.imeraj@sussex.ac.uk}} \\
}
\date{\today\\\small \ This replaces an earlier version of this paper entitled  ``Inverse Options in a Black-Scholes World" }

\begin{document}
	
	\maketitle
	\thispagestyle{empty}
	
	\begin{abstract}
\noindent  Over 90\% of exchange trading on crypto options has always been on the Deribit platform. This centralised crypto exchange only lists inverse products because they do not accept fiat currency. Currently, fiat-based traders can only make deposits in bitcoin, although they can withdraw both bitcoin and ether to their on-chain wallets. Likewise, other major crypto options platforms only list crypto-stablecoin trading pairs in so-called \textit{direct} options, which are similar to the standard crypto options listed by the CME except the U.S. dollar is replaced by a stablecoin version. Until now a  clear mathematical exposition of these products has been lacking. We discuss the sources of market incompleteness in direct and inverse options and compare their pricing and hedging characteristics. Then we discuss the useful applications of currency protected ``quanto" direct and inverse options for fiat-based traders and describe their pricing and hedging characteristics, all in the Black-Scholes setting.
	\end{abstract}

\bigskip



\thispagestyle{empty}

\newpage
\onehalfspacing
\setcounter{page}{1}

\section{Introduction}
Blockchain technology, on which every crypto is built, is rena\^itre.  The storage of information on a distributed ledger which is visible to all participants goes back to the 1990s at least \citep{HS1991,HSS1997}. But the idea of a cryptographically-linked blockchain technology only grew in popularity after the introduction of a crypto currency call bitcoin  on a blockchain called Bitcoin \citep{Satoshi2008}. Bitcoin transactions are verified by a peer-to-peer network, stored on the blockchain, transparent to anyone with  internet access and become immutable after a certain period of time. Today, blockchains of many different types have exceeded their initial purpose of recording simple peer-to-peer transactions, becoming the backbone of Web 3.0 by carrying fully autonomous and self-regulating smart contracts.  As a result, there are tens of thousands of tokens which are transacted on blockchains and also off-chain on market places and centralised exchanges.

A token is a crypto asset that is transacted using a smart contract on a blockchain. Almost all crypto assets are tokens, and even those that were minted before the introduction of Ethereum Request for Comments (ERC) token standards are easily `wrapped' to become a token. Native tokens are generated through building the blocks in the chain. If the blockchain can carry smart contracts the native token is also the unit of account for the `gas' required to fuel the smart-contract transactions.\footnote{For example, ether, the native token of Ethereum is the unit of account for all non-fungible tokens minted onto Ethereum. And although Bitcoin is not a smart contract blockchain, its native token bitcoin is a common crypto unit of account for new token offerings.} Because a native token is used for payment, we can call it a cryptocurrency. Although the term `cryptocurrency' has slipped into popular use as a generic term for all crypto assets, the only other type of token which is  truly a cryptocurrency, i.e. a fungible token that is commonly used for payment and settlements, is a stablecoin. A stablecoin is a token whose price is pegged to a fiat currency such as the U.S. dollar. 

Other tokens which are not used for payment can be divided into fungible and non-fungible classes. A non-fungible token (NFT) is a certificate of ownership of a physical or digital asset such as land or art. NFTs are frequently used as collateral for borrowing cryptocurrencies, but of course they themselves could not be a unit of account because they are non-fungible. An NFT is  an asset which is arguably more akin to a commodity than a security. How to classify the non-payment but still fungible  tokens that are issued by developers of projects in the digital economy is an even greater point of debate. Under the leadership of Gary Gensler and Dan Berkowitz, the U.S. Securities and Exchange Commission (SEC) has been arguing for years that such tokens are a type of security, and should therefore be regulated by the SEC. On the other hand,  the Commodities and Futures Trading Commission (CFTC) produced a counter-argument that the CFTC should be regulating the crypto assets (such as native tokens) that have a utility value and are therefore more like a commodity than a security.

Crypto assets are exchanged in trading pairs via decentralised liquidity pools on-chain or on centralised exchanges off-chain. When both sides of the trading pair are cryptocurrencies, the trading pair is like an exchange rate. If neither side of the trading pair is a cryptocurrency, then the trading pair is more like an asset swap. And if only one side of the trading pair is a cryptocurrency, it is like a security with the cryptocurrency being the unit of account. The most common type of platform for trading exchange-rate crypto pairs is the order book of an off-chain centralised exchange such as Binance or FTX. By contrast, most  asset-swap crypto pairs are traded in liquidity pools of an on-chain  decentralised exchange, such as Uniswap or Sushiswap. Most crypto-crypto asset swaps are  traded in on-chain liquidity pools.  Trading volumes in all these markets have exploded over recent years. In particular, millions of retail traders, either truly believing in a projects' philosophy or for purely speculative purposes, have easily opened accounts on any number of so-called `self-regulated' centralised exchanges to trade  bitcoin and other crypto assets and  derivatives. Large proprietary trading firms have also been trading very actively.\footnote{Recently, BlackRock, Fidelity and Charles Schwab started offering their clients investment opportunities in bitcoin, see \href{https://www.ft.com/content/3261f919-ca98-41d2-b950-bc3a670f994c}{FT.com}. Top tier banks like Goldman Sachs or J.P. Morgan and proprietary trading houses  such as Jump and Cumberland Capital have been active in crypto spot and derivatives market for many years.} All these traders, on entering this melting pot of traditional finance  and modern computer science, are  faced with an array of innovative crypto assets many of which have very actively traded derivative products. However, in the crypto derivatives category, options are backed by very little  academic research.

Almost every crypto derivative has a settlement price based on a non-traded asset. This is because of the extreme fragmentation of the crypto markets, where  trading pairs are typically listed on several different centralised exchanges or traded in a number of competing liquidity pools. For instance,  the bitcoin U.S. dollar trading pair is listed on over twenty exchanges, and pairs of BTC against USD stablecoins are listed on even more \citep{CC_CCCAGG}. For this reason, the settlement price for BTC/USD futures and options is derived from a \textit{coin index} whose value is typically an average price taken across several different exchanges. The exact formula used for the settlement index depends on the exchange. Some also use the average of an index as the settlement price, for instance taken over the last hour or 30 minutes before  settlement. So technically, these crypto derivatives markets are incomplete.

At the time of writing two-thirds of total trading volumes on all centralised crypto exchanges is on derivative products rather than spot trading pairs, see \cite{CC_August} . These products include standard, direct and inverse calendar futures and options, and perpetual futures, the latter  presenting by far the greatest trading volume in the entire crypto asset ecosystem. Perpetual contracts  are non-expiring futures that mimic a margin-based spot account on the coin index.  A regular funding payment assures the agent that the price of the perpetual stays close to the underlying coin index value, hence these products are often called perpetual swaps. While a perpetual contract is traded and never settled, its price is not the same as the settlement price for the corresponding futures and options because the basis risk can be high on these highly volatile instruments.   

More recently, a great many centralised exchanges have also listed European-style options, mainly on BTC and ether (ETH) against USD. Some of these platforms are at least semi-regulated by the CFTC, including the Chicago Mercantile Exchange (CME), FTX.US (formerly LedgerX) and IQ Option. Others, such as Deribit are  registered in off-shore tax havens where there is no need to comply with know-your-customer protocols, market abuse directives or indeed any other form of regulation aimed at protecting the client's interests. Due to the  rapid growth in  popularity of crypto options, by May 2021  Deribit's monthly trading volume of bitcoin options was \$26bn and on ether options it was \$15.5bn. At the time of writing, just after the merge of the Ethereum chain to proof-of-stake consensus, ETH options have a greater volume than BTC options on most exchanges.\footnote{For instance, for the month of August 2022 Deribit's ETH options volume was \$11.7bn compared with their BTC options volume of \$9.26bn, see \href{https://analytics.skew.com/dashboard/ether-options}{skew.com} for detailed volume data.}.  

Apart from regulatory oversight, the main factor that differentiates the centralised crypto derivatives  platforms is the type of products that they offer. For example,  CME and FTX.US run order books in standard European put and calls with a contract size in bitcoin or ether, and all contracts are margined and settled in USD.\footnote{See, for instance, \href{https://derivs.ftx.us/options}{FTX options contract specifications}. The CME lists two types of options contracts: standard and micro, i.e. an option on one or one-tenth of a bitcoin. FTX offers a wide range of crypto options through a \emph{Request for Quote} system and IQ Option offers several products, including binary options,  but only to professional traders. }  But $\sim 90\%$ of open interest and trading volume on crypto options has always been on Deribit, which only runs order books in so-called \textit{inverse} options. These have a contract size of one bitcoin, ether or sol, and they track the USD value of these coins even though they are margined and settled in BTC, ETH or SOL. Such an inverse structure is necessary because clients cannot use fiat currency for on-boarding or off-boarding the Deribit exchange, even though they wish to trade an option on a cryptocurrency-USD pair. In fact, at the time of writing, Derbit \textit{only} allows deposits in BTC,\footnote{BTC  can then be swapped for ETH and/or SOL and/or the stablecoin USDC using the \href{https://www.deribit.com/kb/swap-service}{Deribit swap service.} No other brokerage-type services are provided.} and withdrawals in BTC or ETH directly to the blockchain wallets. This means that any USD-denominated agent trading inverse options on Deribit must bear the expense of first transferring the profit in BTC or ETH to their wallet, and then sending it to another crypto exchange, where it can eventually be converted to USD. Other large but self-regulated -- let us call them unregulated -- exchanges also prevent fiat off-boarding but they do allow two-way transfers of stablecoins such as USDT or BUSD. This allows them to list \textit{direct} options which are exactly like the CME and FTX.US products, except that margining and settlement is in a stablecoin. Either way, the absence of on-exchange crypto-fiat brokerage services  makes crypto options trading a cost-inefficient operation for USD-denominated traders.

This discussion has highlighted three important questions to answer before we can really understand crypto options should be priced and hedged:
\begin{enumerate}
 \item What is the currency that the trader uses as a unit of account? This could  be a cryptocurrency like BTC or a fiat currency such as USD. A BTC-based trader with a BTC-denominated trading account has a different perspective on profit and loss to a US trader whose account is measured in USD; 
    \item Is the underlying a security or a currency?\footnote{Or indeed an asset swap. Crypto-crypto asset swaps are most heavily traded in decentralised liquidity pools, recorded on blockchains, i.e. `on-chain'. Not dissimilar to over-the-counter (OTC) agreements in traditional markets, except that on-chain transactions are  fully transparent. But block production is rather slow so it could be a relatively long time before crypto-crypto on-chain swap derivatives are developed.} If a  security, then the option should be priced like a stock or bond option  with the unit of account specified in 1, and if a currency then the option is equivalent to an FX option;
    \item Is the settlement price that of a tradable instrument? The vast majority of crypto options are settled using an index, sometimes also averaged prior to settlement. So should pricing and hedging take account of the incompleteness of this market or not?
    \end{enumerate}
Clarifying the answers to those questions leads us to suggest that \textit{quanto}  options be added to the array of crypto products. Quanto direct options are similar to traditional quantos, but the quanto inverse option is a completely new type of exotic option. We argue that both direct and inverse quantos are better products for risk-averse USD-denominated agents than their vanilla counterparts which have no currency protection. 

In the following: Section \ref{sec:products} clarifies the exact differences between standard, direct and inverse options, discusses the applications of quanto direct options for crypto traders and introduces quanto inverse options; Section \ref{sec:pricing} derives pricing formulae for  inverse and quanto inverse options under the geometric Brownian motion assumption and examines the associated issue of market incompleteness; Section \ref{sec:greeks} derives the Black-Scholes type hedge ratios for quanto inverse options and describes their characteristics; and Section \ref{sec:conc} concludes.

\section{Crypto Option Types}\label{sec:products}
We now define the payoffs and settlements for direct and inverse options precisely using USD as the fiat side and either bitcoin (BTC) or ether (ETH) as the crypto side of the trading pair. Both bitcoin and ether are cryptocurrencies, but bitcoin is only a cryptocurrency while ether can also be regarded as a security or commodity because Ethereum is a smart contract blockchain whereas Bitcoin is not. Indeed, the SEC have argued that \textit{every} token apart from bitcoin is a security and should therefore fall under the jurisdiction of the SEC rather than the CFTC.\footnote{ {In a recent \href{https://www.sec.gov/news/speech/gensler-sec-speaks-090822}{SEC speech}, Gary Gensler points out that ``\textit{the vast majority [of crypto tokens] are securities}'' but excludes bitcoin in particular in an earlier \href{https://www.cnbc.com/video/2022/06/27/sec-chair-gary-gensler-discusses-potential-crypto-regulation-and-stablecoins.html?&qsearchterm=gary20gensler}{interview}. The argument is that native token of other blockchains that are not smart-contract compliant, such as Dogecoin can be thought of a security because their primary use is as a `meme' token. Just recently, the \href{https://uk.finance.yahoo.com/news/ethereum-price-drops-sec-declares-control-093528955.html}{SEC announced} to classify Ethereum and all its subsidiary tokens as a security. Thus, every project deployed on top of the Ethereum blockchain could be claimed to be a security and within SEC jurisdiction.}} We also consider two types of traders, one USD-denominated whose trading book is denominated in USD and the other crypto-based whose trading book is denominated in BTC.

Both settlement mechanisms for crypto options which differ between exchanges as well as the different types of products are not yet widely understood. For instance, on  Deribit  the underlying  is a non-tradable  index of spot prices and on the CME it is a futures contract on a similarly non-tradable reference rate. The settlement price on Deribit is the average value of the underlying over the 30 minutes prior to settlement and on the CME it is the spot value of the reference rate.\footnote{But the public information on Derbit employs confusing terminology,  specifying neither how the index is weighted  nor how frequently the underling price is monitored during the 30-minute interval. For example: ``\textit{Exercise of an options contract will result in a settlement in BTC immediately after the expiry. The exercise-settlement value is calculated using the average of the Deribit BTC index over the last 30 minutes before the expiry. The settlement amount in USD is equal to the difference between the exercise value and the strike price of the option. The exercise value is the 30 min average of the BTC index as calculated before the expiry. The settlement amount in BTC is calculated by dividing this difference by the exercise value.}". They have hardly any market share compared with Deribit, but the CME Group are crystal clear: ``\textit{The underlying for CME options on Bitcoin futures is one CME Bitcoin futures contract. As you know, The CME Bitcoin futures contract represents five bitcoin and cash settles to the CME CF Bitcoin Reference Rate (BRR).}" See \href{https://www.cmegroup.com/education/courses/introduction-to-bitcoin/get-to-know-options-on-bitcoin-futures.html}{CME Group}. } These settlement differences, combined with a widespread lack of proper documentation from the unregulated exchanges, may lead to confusion about a seemingly trivial European-style product.

We begin by comparing standard options with direct options and inverse (sometimes also called indirect) options, then we discuss the uses of  quanto direct options, as well as a new type of quanto inverse option. It is quite possible that quanto direct and/or inverse options are already being traded on-chain, just as traditional quanto options are traded over-the-counter, and so our first through specification of details in this section could be very useful to their users.

\subsection{Standard vs Direct Options}\label{sec:direct}
Either standard or direct options are widely traded on all the regulated and some unregulated centralised exchanges. A bitcoin (ether) option trader can choose whether to trade the pair BTC-USD (ETH-USD)  or BTC-USDT (ETH-USDT) or the other side of the pair could be some other stablecoin such as USDC. Because of the risks surrounding stablecoins,\footnote{The risks are huge. They include:  market risk (because the stablecoin is only pegged to the value of USD and can deviate very far from the peg, in May 2022); but also operational risks (stablecoins are held on blockchains and so can be hacked); and regulatory risks (e.g. the European Markets in Crypto Assets (MiCA) directive places very firm caps on stablecoin trading volumes, to try to limit their capitalisation).} a USD-denominated trader may prefer the standard option, which is a plain vanilla European product. The call has payoff in USD given by:
\begin{equation}\label{eq:standard_payoff}
	V_{_T}^{^{_{\$}}} =  \left(S_{_T}^{^{_{\$}}} - K^{^{_{\$}}} \right)^+,
\end{equation}
where $K^{^{_{\$}}}$ denotes the strike price and $S_{_T}^{^{_{\$}}}$ denotes the underlying price at maturity $T$. On the CME (at the time of writing) the underlying is either the BTC-USD pair -- which from now on we regard as the BTC/USD, exchange rate, i.e. the number of USD for 1 unit of bitcoin\footnote{This follows financial market convention, although economists and physicists would naturally regard the notation $\mathbbm{Y/Z}$ to mean the number of units of $\mathbbm{Y}$ per unit of $\mathbbm{Z}$. But for currency quotes it is the other way around.} --  or the pair ETH-USD which we regard as either the ETH/USD exchange rate or the USD price of the ETH security token. 

The margining and settlement of standard bitcoin or ether options depends on the exchange. On the CME they are margined and settled in USD and the settlement price is that of the CME bitcoin or ether futures, so both the option premium and the payoff are  denominated in USD. They are also cleared by the exchange, thus omitting counterparty risk. This procedure does not differ in any way from trading other commodities on the CME exchange. 

On FTX.US, the underlying is bitcoin (or ether) and the settlement price is the  BTC/USD (or ETH/USD) price on the FTX.US spot platform at the time of settlement. The FTX.US settlement style differs from the CME and so does the collateral structure because it depends on the option type, being designed to eliminate any counterparty risk.\footnote{A long call is \textit{physically} settled. A bitcoin options trader purchasing a call pays the premium in USD and, at maturity, has the right to buy a bitcoin for $K^{^{_{\$}}}$, hence receiving a BTC-denominated payoff. By contrast, on taking a short  call position the trader receives the premium in USD but is required to hold a long position in the underlying, i.e. only covered calls are allowed. The terminal payoff of a short call is therefore cash-settled in USD. Similarly, to purchase a long position in a put, the agent pays a USD-denominated premium and is required to hold the notional option amount as collateral (protective put). A short put position is achieved by a cash-covered put position, i.e. the agent deposits $K^{^{_{\$}}}$ and receives the premium in USD. See \href{https://support.ledgerx.com/hc/en-us/articles/360044339734-How-to-Place-an-Options-Trade-on-FTX-US-Derivatives}{FTX Options Trading Specification} for more details on buying/selling bitcoin options on FTX.US and  \href{https://support.ledgerx.com/hc/en-us/articles/4402493969171-Options-Collateral}{FTX Options Collateral} for detailed collateral requirements. }

Binance lists \textit{direct} BTC-USDT and ETH-USDT (and many other crypto) options which are margined and settled in USDT. The settlement price is that of the stablecoin-margined product BTCUSDT (or ETHUSDT) on their spot platform at the time of settlement, and the payoff is in USDT;  but in other ways  the Binance settlement procedure mimics the CME's. Hence, using  the symbol $\mathbbm{T}$ to denote the price of the stablecoin USDT, the call pay-off may be written: 
\begin{equation}\label{eq:direct_payoff}
	V_{_T}^{^{_{\mathbbm{T}}}} =  \left(S_{_T}^{^{_{\mathbbm{T}}}} - K^{^{_{\mathbbm{T}}}} \right)^+,
\end{equation}
which is identical to \eqref{eq:standard_payoff} except the currency is not but the stablecoin USDT.

\begin{figure}[h!]
	\centering
	\small
	\caption{Standard and Direct Option Payoff}
	\vspace{-0.3cm}
	\caption*{\footnotesize The payoff to a long call (left; blue graph) and put (right; red graph) as a function of the settlement price. These payoffs present a standard option but could also display the direct payoff using USDT ($\mathbbm{T}$) instead of USD (\$) for both the payoff (vertical axis) as well as the underlying (vertical axis). The strike level is set at $K^{^{_{\$}}}$ = \$25,000. }  
	\includegraphics[width=0.49\textwidth]{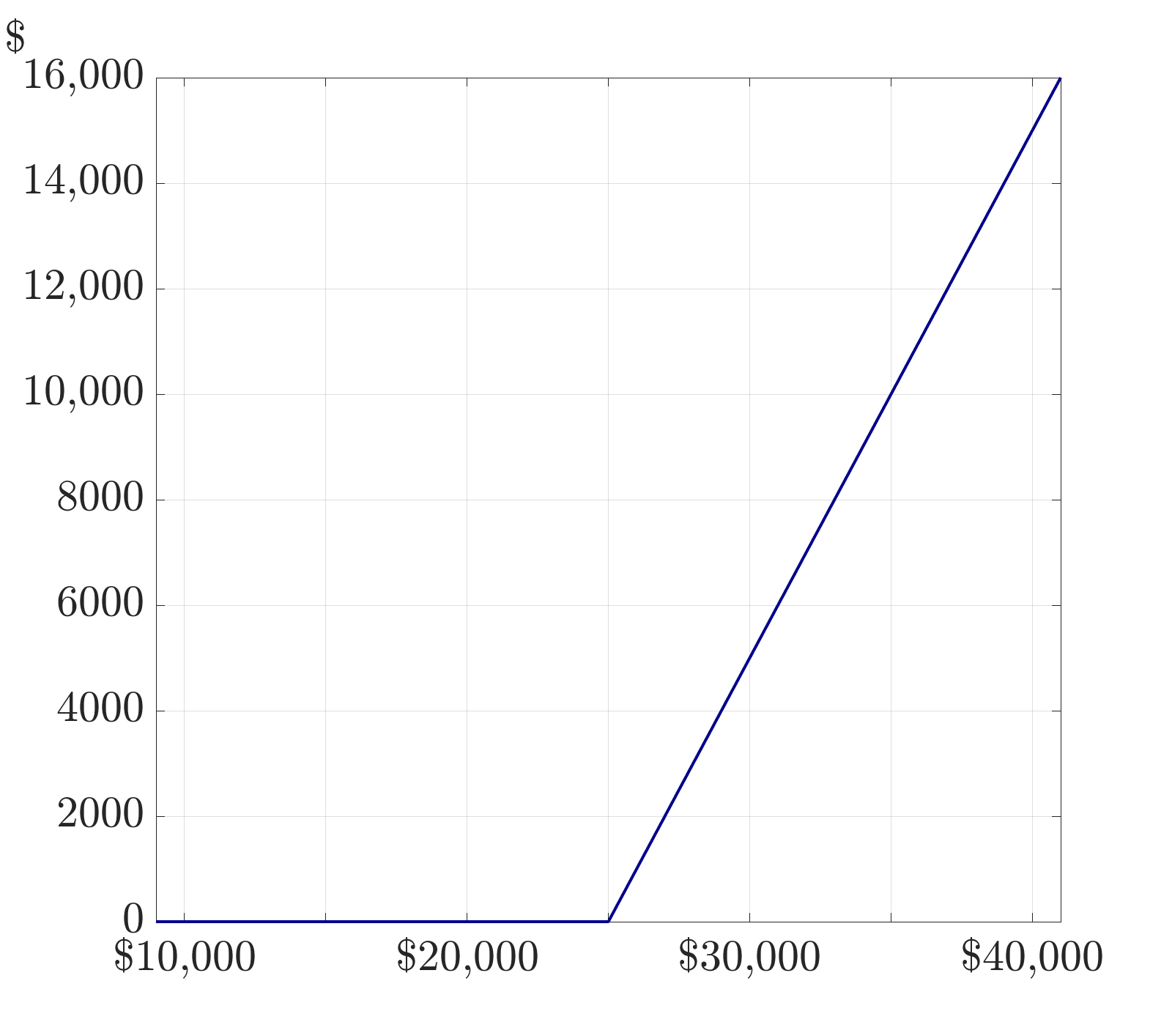}
	\includegraphics[width=0.49\textwidth]{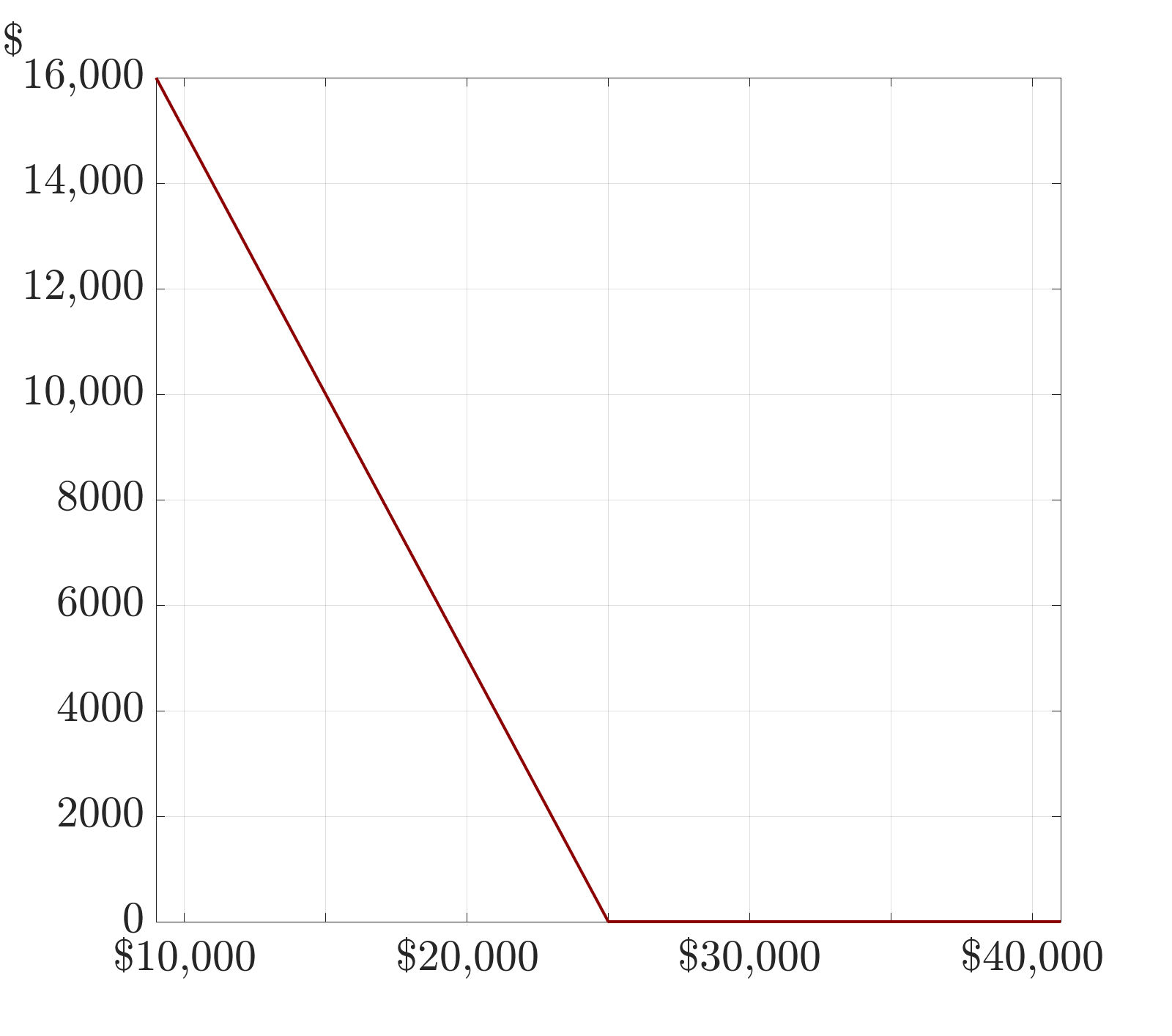}	
	\label{fig:direct_standard_Example} 
	\vspace{-12pt}
\end{figure}

We have selected these three particular centralised exchanges because they are the largest to list standard and direct crypto options, and they illustrate how three exchanges can offer the same basic product, with but have different underlyings and settlement procedures.\footnote{ On the other side of the spectrum, more decentralised options exchanged are emerging. For instance, Ribbon Finance is a decentralised cryptocurrency option exchange built on the Ethereum blockchain. It allows traders to lock their investments into ``Theta Vaults'', sometimes called ``DeFi Option Vaults'', and generate weekly yields. These vaults run various automatic option strategies, e.g. covered calls or short puts, through minting and shorting \href{https://www.opyn.co/}{Opyn's} oToken. The oTokens are effectively cash-settled option contracts, as they give the holder the right to redeem an amount of the underlying if the the smart option contract ends up ITM. The contract specifications as well as the premium is set via auction, see \href{https://docs.ribbon.finance/theta-vault/theta-vault}{Theta Vaults}. Other protocols offering DeFi cryptocurrency options exposure are \href{https://www.thetanuts.finance/}{Thetanuts} or Opyn's own \href{https://squeeth.opyn.co/?ct=GB}{Squeeth}. } 
We also note that, despite being the only crypto options exchange not to run 24/7 on every day of the year, currently the CME has much the greatest trading volume and open interest of the three. This could be because of the dominance of USD-denominated traders, who prefer to use USD rather than a stablecoin as the unit of account. Figure \ref{fig:direct_standard_Example} illustrates the well-known payoffs of standard and direct call and puts as a function of the underlying. For the sake of clarity, we omit the graphs for USDT as these would not add any further information.

\subsection{Inverse Options}\label{sec:inverse}
Inverse options are the only kind of option Deribit lists, and as such they are responsible for over 90\% of the trading volume on centralised crypto options exchanges. Deribit is a non-fiat exchange -- so there is no USD transacted anywhere on the platform -- nevertheless Deribit lists options on three cryptocurrency-USD trading pairs, i.e. BTC-USD, ETH-USD and just recently SOL-USD, all options being margined and settled  in the cryptocurrency of the underlying option. 

Many other non-fiat exchanges list inverse options, precisely because the can trade against the USD without using it as the unit of account or indeed allowing any fiat currency onto the platform. And most inverse options  track the USD value of a coin \textit{index} not a single spot price. Importantly, margining and settlement is always in a cryptocurrency, not in fiat. If held to maturity, the settlement price $S_{_T}$ is either the coin index value exactly at the settlement time, or its average value over a time period immediately prior to settlement. For example, Deribit bitcoin inverse options use the `Deribit Bitcoin Index' for settlement. This is currently  an average of bitcoin spot prices on eleven major centralised exchanges. Any constituent value that falls outside the +/-0.5\% range of the median price is adjusted to the closest bandwidth price limit, and then the index is calculated as the equally-weighted average of these values. The exchange also reserves the right to manually exclude exchanges from this index, if it sees fit.\footnote{ See \href{https://www.deribit.com/main/indexes}{Deribit Index} for a more detailed explanation.} 

Due to its frequent rebalancing, physical replication of this index is an immensely difficult and expensive task. An agent would be required to hold bitcoin positions on multiple exchanges and rebalance these constantly. Complicating replication, even more, the final option settlement value is the average of the index during the last 30 minutes before expiry. This important feature about Deribit inverse options is often left out. The underlying is not directly tradable. For this, we must consider pricing within an incomplete market.

In general, an inverse contract specifies a notional number $N$ of coins  which is multiplied by a point value to obtain a payoff expressed as a number of coins, i.e. in the units of the coin. The terminal payoff (and indeed all trading profits) are transferred to the trader in the cryptocurrency, not in USD. For example,  the payoff $V^{^{_{\mathbb{B}}}}_{_T}$ to an inverse call on BTC/USD is denominated in BTC and may be written:
    \begin{equation}\label{eq:inverse_payoff}
	V^{^{_{\mathbb{B}}}}_{_T} =  N\, \frac{\left(S_{_T}^{^{_{\$}}} - K^{^{_{\$}}} \right)^+}{S_{_T}^{^{_{\$}}}},
    \end{equation} 
  where the second term is a dimensionless quantity called the point value. There is actually no need for our notation to specify the units for the settlement price and strike (even though we have done so above)  because their difference becomes dimensionless when divided by $S_{_T}^{^{_{\$}}}$. But we do need to know what the underlying is and in this example it is the BTC/USD exchange rate. 
  
 Currently, all exchanges that list inverse  options use a notional of exactly one coin,\footnote{Deribit set $N=1$ bitcoin, or 1 ether or 1 sol, depending on the underlying. See \href{https://www.deribit.com/kb/options}{Deribit Option Specification} for example.} Moreover, they all quote the option prices in USD, as well as in the cryptocurrency of the trading pair, but settlement is always in the cryptocurrency. Now, for a USD-denominated trader the true payoff (which is in cryptocurrency) may just as well be translated into USD, in which case   $V^{^{_{\mathbb{B}}}}_{_T}$ should be multiplied by the price of the underling at the time of settlement $\Bar{S}_{_T}^{^{_{\$}}}$. Note that this price is not the same as the settlement price, the latter being the average price over the 30 minutes before the settlement time. Given how volatile crypto markets are, there can actually be a large difference. Anyway, we can write the payoff to a USD-denominated trader as:
   \begin{equation}\label{eq:inverse_payoff_USD}
	V^{^{_{\mathbb{\$}}}}_{_T} = \Bar{S}_{_T}^{^{_{\$}}} V^{^{_{\mathbb{B}}}}_{_T} =\Bar{S}_{_T}^{^{_{\$}}} \frac{\left(S_{_T}^{^{_{\$}}} - K ^{^{_{\$}}}\right)^+}{S_{_T}^{^{_{\$}}}}\approx \left(S_{_T}^{^{_{\$}}} - K^{^{_{\$}}} \right)^+.
    \end{equation} 
This shows that one can think of an inverse option as a standard FX option except the payoff is denominated in the foreign currency -- as already remarked by \cite{Lucic2022}. There is a large body of academic research on  FX options, their pricing, hedging, volatility dynamics and so forth, see \cite{Levy1992}, \cite{CW2007}, and \cite{Demeterfi1998} and many others. However, FX options are usually denominated in the same currency as the underlying. Inverse options are denominated in the foreign rather than the domestic currency and this rather unusual denomination of the payoff is a potential source of confusion. Because the payoff to a long inverse call has an upper bound of one bitcoin, the option's delta at maturity is zero up to the strike price, jumps up to one beyond this price -- just like a standard call -- but then it decreases, eventually approaching zero as $S_{_T} \rightarrow \infty$. We shall illustrate this later in Figure \ref{fig:Greeks1Call} and \ref{fig:Greeks1Put}. Similar characteristics in FX options, when one uses the foreign-domestic symmetry relationship to convert a domestic call to a foreign put, has been documented previously \citep{Grabbe1983, RW2010}. We also provide more details on this relationship,  in Section \ref{eq:Inverse_Call_Pricing}. 

\begin{figure}[h!]
	\centering
	\small
	\caption{Inverse Option Payoff}
	\vspace{-0.3cm}
	\caption*{\footnotesize The payoff in bitcoin to a long inverse call (left; blue graph) and put (right; red graph)  as a function of the settlement price.  The strike is \$25,000. Note that the call payoff is capped at maximum $\mathbbm{B}1$ whereas the put can theoretically pay out an infinite amount of bitcoin. }  
	\includegraphics[width=0.49\textwidth]{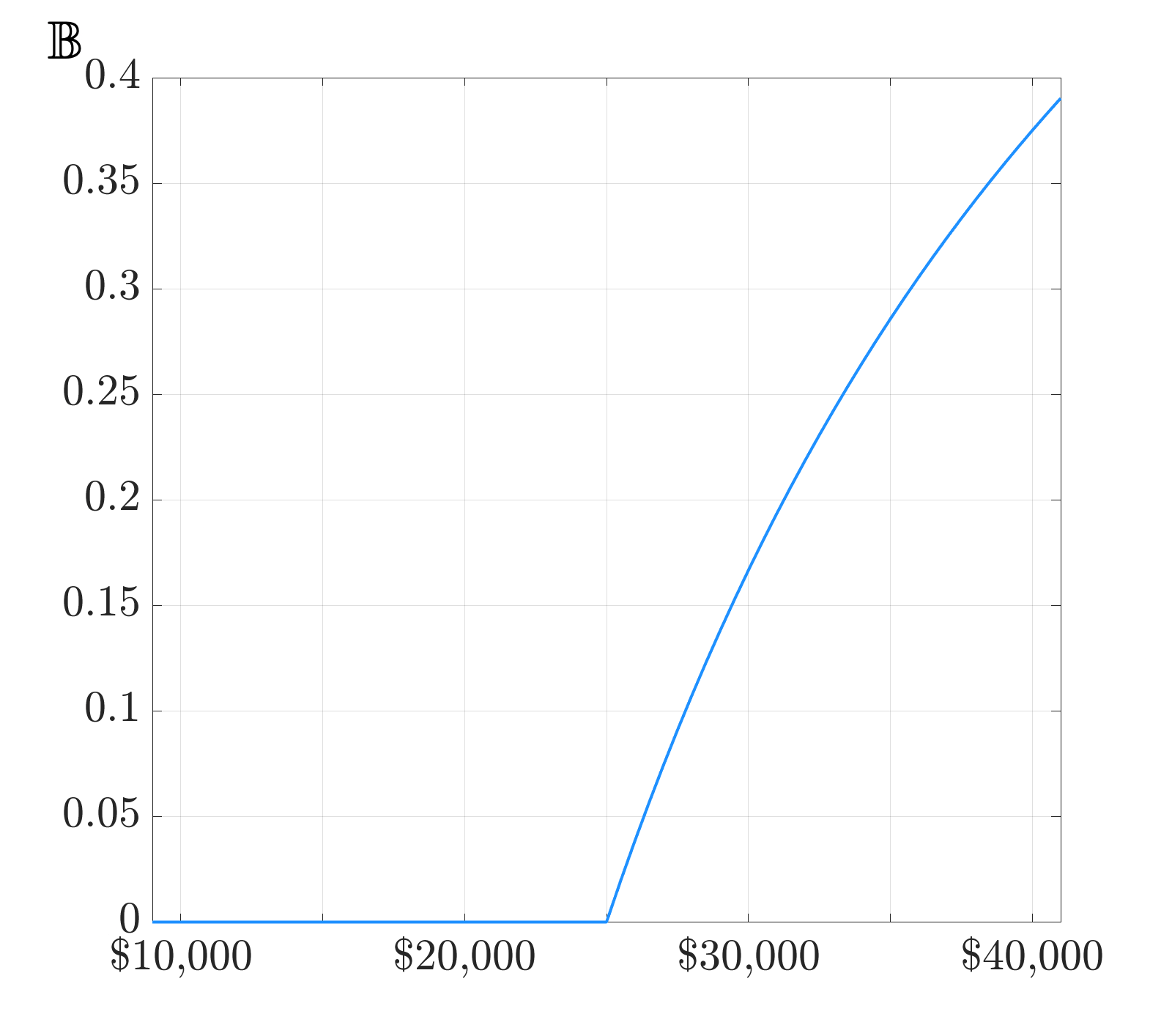}
	\includegraphics[width=0.49\textwidth]{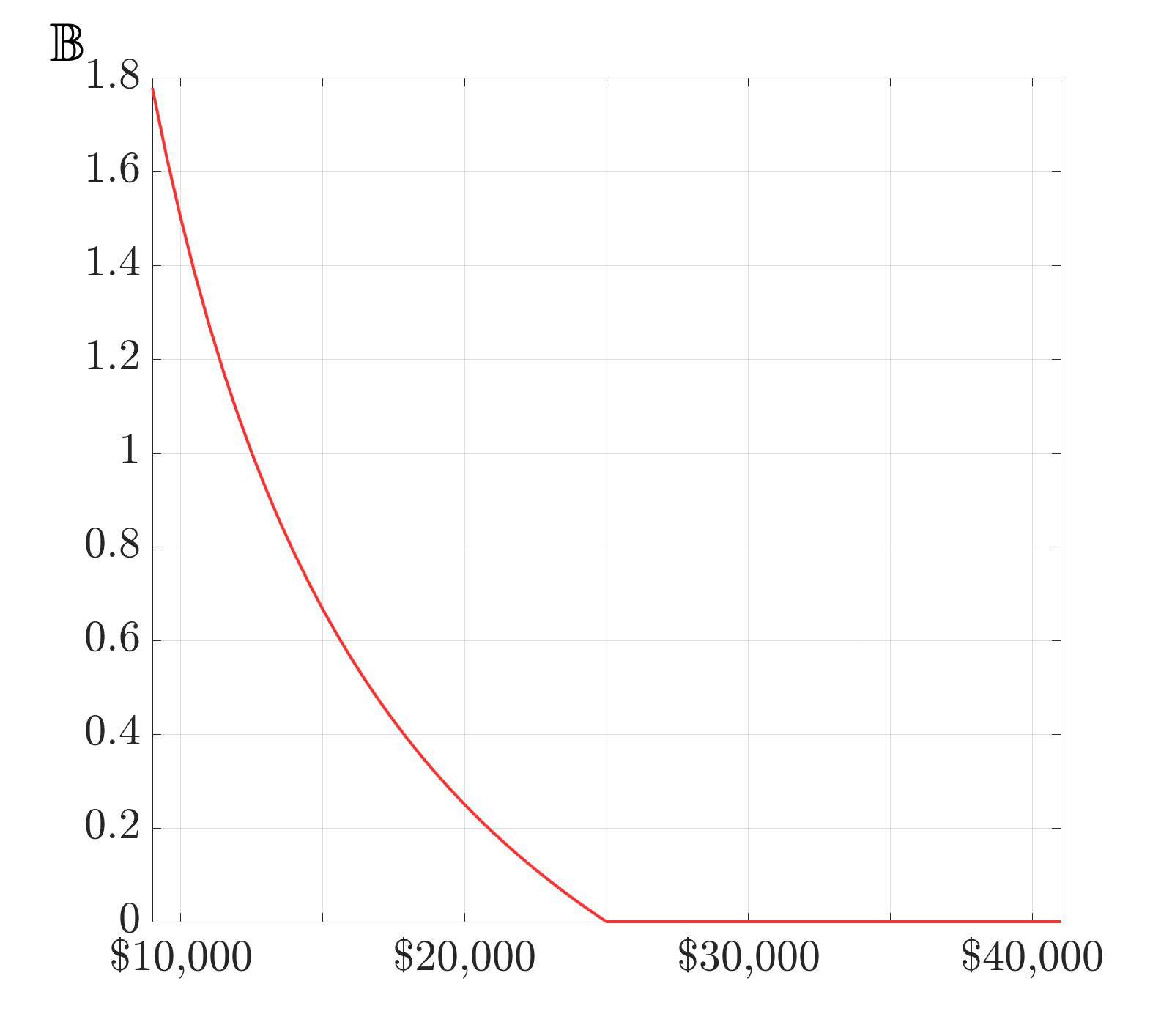}	
	\label{fig:inverse_Example} 
	\vspace{-12pt}
\end{figure}

 Figure \ref{fig:inverse_Example} illustrates the true terminal payoff an agent would receive when trading a long inverse option. It is a piece-wise  concave (call) or entirely convex (put) structure. Note that the call payoff is capped at maximum $\mathbbm{B}1$ whereas the put can theoretically pay out an infinite amount of bitcoin. For USD-denominated traders, but only for these traders, the payoff to an inverse option  can be approximated by a standard piece-wise linear form,\footnote{Approximate because $S^*_{_T}\ne S_{_T}$, unless there is no trading in the 30 minutes before expiry. However, this difference affects only the payoff and not the fair price at any time prior to expiry.} and hence their pricing is almost identical to that of the direct option \citep{GK1983}. But very much  depends only on the trader's base currency.  USD-denominated traders might only consider the USD value of their option position on the balance sheet, but crypto option traders are international, and so may prefer to use a cryptocurrency -- or a different fiat currency -- as their unit of account on Deribit. And even for USD-denominated traders, cryptocurrency-denominated profits stemming from large positions in inverse options cannot be instantly exchanged for USD. This is a particular issue for puts which can, in theory, pay out an infinite amount of bitcoin. Deribit also offers block trades of at least 25 bitcoin or 250 ether  for which it does not require very large negative price movements for the position holder to receive a high number of bitcoin or ether. Thus, while the \textit{paper} value of  profits in BTC or ETH may be very high, converting such an amount into fiat may well result in liquidity problems. Then profits would be reduced through spillage on the currency conversion trades. The absence of adequate brokerage services, and of exchange requirements to hold large margin reserves, are yet another expense factor for institutional bitcoin option trading. Professional traders transacting large amounts might refrain from denominating their profit and loss in USD and use cryptocurrency their balance sheets instead, but still the currency risk faced  traders in inverse options cannot be ignored.
 
\subsection{Standard Quanto and Direct Options}\label{sec:quanto}
A quantity-adjusted option, or quanto option for short,  allows an trader to gain exposure to a foreign market without taking any currency risk. In traditional markets the underlying is often a single security, or a security index, or another asset like a commodity. The settlement price $S_{_T}$ of this asset and the option strike  $K$ are denominated in the foreign currency but the standard option payoff is converted into domestic currency using a predetermined exchange rate $\Bar{X}$ which is agreed upon entering the contract (quanto futures and options are usually traded OTC). For instance, a standard quanto call has payoff:
\begin{equation}\label{eq:quanto_vanilla_payoff}
	V_{_T} =  \Bar{X} \,\left(S_{_T} - K \right)^+.
\end{equation}
Because they are well-known products in traditional markets the pricing and hedging for quanto options has been researched very extensively \citep{Jamshidian1993, Demeterfi1998, JYC2009, Clark2011}.  So it is somewhat surprising that there has been no previous research documenting quanto products in  crypto markets. In fact, the first futures ever traded on crypto were quanto products.\footnote{The derivatives exchange BitMEX was among the first exchanges to offer quanto products in form of futures on various coin/coin or coin/USDT pairs, see \href{https://www.bitmex.com/app/quantoFuturesGuide}{BitMEX quanto futures}.} 

To illustrate the usefulness of a quanto option to crypto traders, first suppose a BTC-based option trader is interested to gain exposure to ETH. She could use existing exchange-traded products to convert BTC to USD and then trade ETH-USD options. But this way she is exposed to the risk of BTC fluctuating against USD. To remove the currency risk the trader can obtain a payoff denominated in BTC by agreeing a fixed USD/BTC rate $\Bar{X}^{^{_{\$/\mathbbm{B}}}}$ with the quanto option issuer before entering the contract.\footnote{Recall that (counter-intuitively) the notation $\mathbbm{Y/Z}$ is the number of units of $\mathbbm{Z}$ per unit of $\mathbbm{Y}$, so multiplication by the quanto factor $\Bar{X}^{^{\mathbbm{Y/Z}}}$ converts the units from $\mathbbm{Y}$ to $\mathbbm{Z}$. } We express the payoff to a quanto call, for a BTC-based trader as:
    \begin{equation}\label{eq:quanto_standard_payoff}
 	V_{_T}^{^{_{\mathbbm{B}}}} = \Bar{X}^{^{_{\$/\mathbbm{B}}}} \left(S_{_T}^{^{_{\$}}} - K^{^{_{\$}}} \right)^+,
    \end{equation}    
where $S_{_T}^{^{_\$}}$ is the price of  ETH in USD   at maturity $T$ and the option strike $K^{^{_{\$}}}$ is also denominated in USD.  These types of options would enable traders to participate in ETH without physically owning or depositing ether, thus avoiding gas or other blockchain fees. 

\begin{figure}[h!]
	\centering
	\small
	\caption{Standard Quanto Option Payoffs}
	\vspace{-0.3cm}
	\caption*{\footnotesize The payoffs to a long standard quanto call (left; blue graph) and put (right, red graph). The security (in this case) is ether, the foreign currency is USD and the domestic currency is BTC, i.e. the buyer is a BTC-based trader. Thus, the payoff is in BTC and the underlying on the horizontal axis is the USD price of ETH. We set  $K^{^{_{\mathbbm{E}}}}$ = \$1,750. }  
	\includegraphics[width=0.49\textwidth]{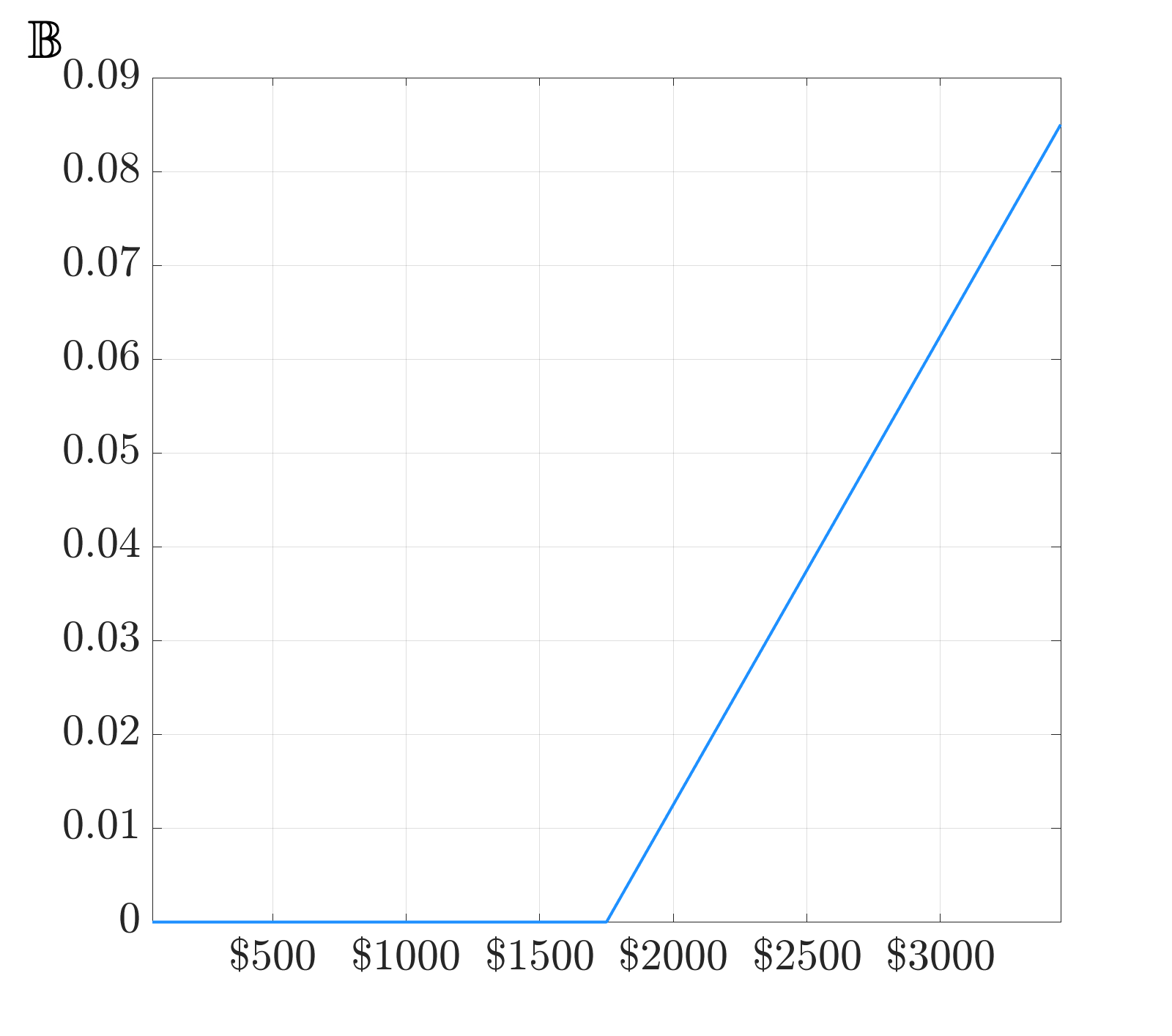}
	\includegraphics[width=0.49\textwidth]{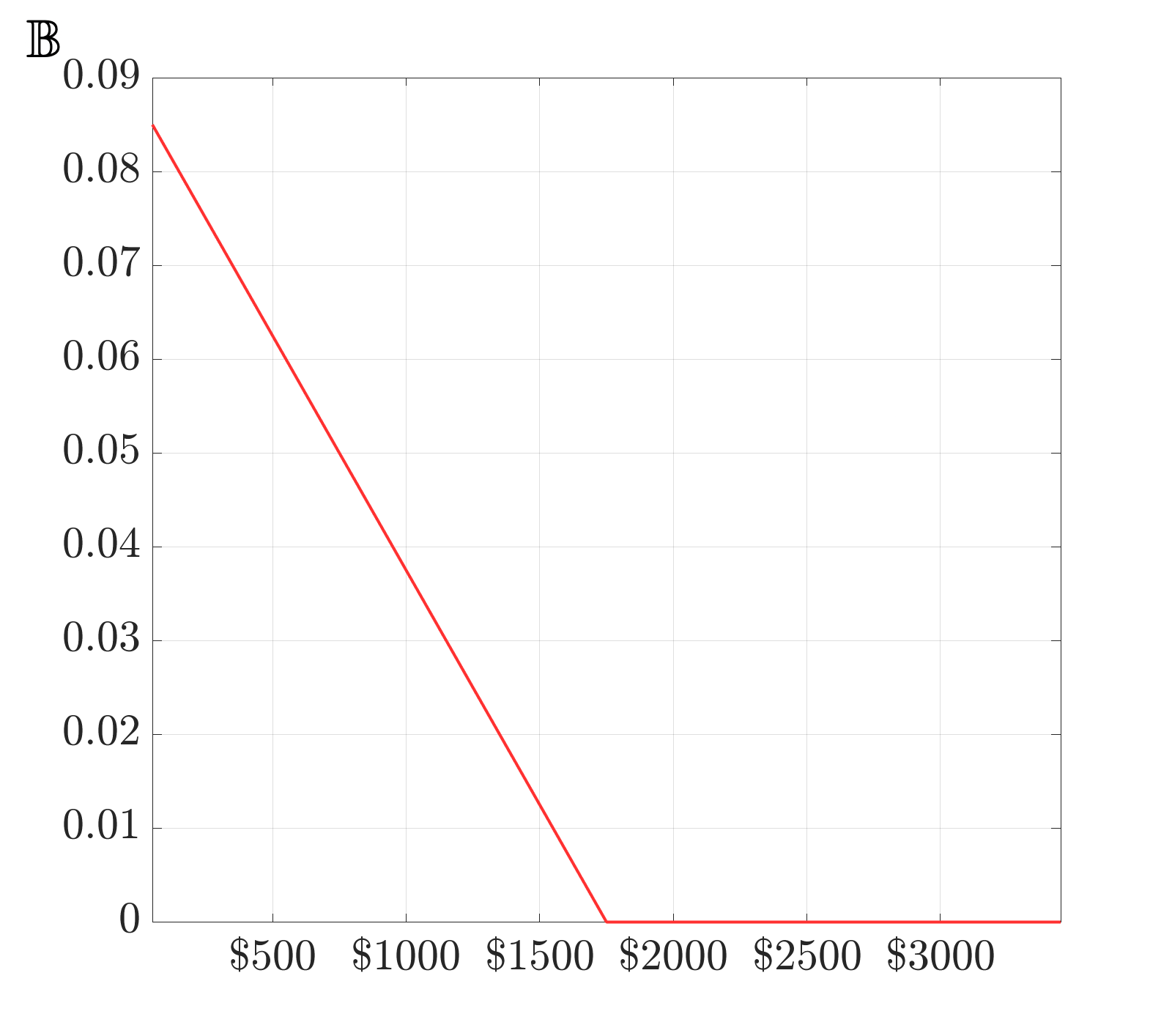}	
	\label{fig:Standard_Quanto_Example} 
	\vspace{-12pt}
\end{figure}

Figure \ref{fig:Standard_Quanto_Example} depicts the payoffs to a quanto call and put  given by (\ref{eq:quanto_standard_payoff}).  Note the difference between inverse  and $\$/\mathbb{B}$ quanto option, i.e. the call quanto payoff is not capped upwards, whereas the inverse pays out a maximum of $\mathbb{B}$1. It is the opposite for the put, the inverse is uncapped and could  theoretically pay out an infinite amount.

The main cryptocurrencies like bitcoin and ether have very active options markets but there are thousands of minor tokens that are \textit{only} paired with stablecoins on the major non-fiat exchanges like Binance, never with USD. Therefore a USD-denominated trader wishing to trade such a token -- let us call it XYZ -- must take on the (very real) risk of the stablecoin de-pegging from USD while the trade on XYZ is in place. To remove this currency risk, at the same time as leveraging their exposure to XYZ though an option trade, a  USD-denominated agent might agree a fixed stablecoin exchange rate  with the quanto option issuer. For instance, using a fix of the tether rate $\Bar{X}^{^{_{\mathbbm{T}/\$}}}$ the quanto \textit{direct} call USD-denominated payoff becomes:
 \begin{equation}\label{eq:quanto_direct_payoff}
 	V_{_T}^{^{_{\$}}} = \Bar{X}^{^{_{\mathbbm{T}/\$}}}  \left(S_{_T}^{^{_{_\mathbbm{T}}}} - K^{^{_{\mathbbm{T}}}} \right)^+,
    \end{equation}    
where $S_{_T}^{^{_\mathbbm{T}}}$ is the tether price of the XYZ token at the time of the option maturity $T$ and the option strike $K^{^{_\mathbbm{T}}}$ of the quanto is also denominated in USDT.

\subsection{Quanto Inverse Options}\label{sec:quantoinverse}
A quanto inverse option is a natural extension of both inverse and quanto direct options which converts the inverse option payoff to another currency using an exchange rate that is fixed upfront and agreed between both parties when the contract is issued. This allows traders to mitigate the currency risk that we discussed in Section \ref{sec:inverse} as an unavoidable concern about inverse options, because all the exchanges listing these options use cryptocurrency as their unit of account. 

The payoff to a quanto inverse call with a notional of $N=1$ coin in cryptocurrency $\mathbbm{Y}$ can be converted to a currency $\mathbbm{Z}$ (either crypto or fiat) at a fixed exchange rate $\Bar{X}^{^{_{\mathbb{Y}/\mathbbm{Z}}}}$, yielding the payoff: 
   \begin{equation}\label{eq:quanto_inverse_payoff}
	V^{^{_{\mathbb{Z}}}}_{_T} = \Bar{X}^{^{_{\mathbb{Y}/\mathbbm{Z}}}}\, \frac{\left(S_{_T}^{^{\mathbbm{Z}}} - K^{^{\mathbbm{Z}}} \right)^+}{S_{_T}^{^{\mathbbm{Z}}}},
    \end{equation}     
where  $S_{_T}^{^{\mathbbm{Z}}}$ and $K^{^{\mathbbm{Z}}}$ are the settlement and strike prices of an option on a token XYZ, both denominated the cryptocurrency $\mathbbm{Z}$  
These options have payoffs that mimic the convex put and concave call payoffs of inverse options, but they are denominated in a different currency.\footnote{ In fact,  the quanto inverse option has  intuition similar to the self-quanto, sometimes also known as a quadratic option, where a payoff is expressed in units of the foreign currency \citep{Mercurio2003}.}  The quanto factor $\Bar{X}^{^{_{\mathbb{Y}/\mathbbm{Z}}}}$ changes the slope of the terminal payoff and consequently also the option price prior to expiry. Depending on the choice of $\Bar{X}^{^{_{\mathbb{Y}/\mathbbm{Z}}}}$, a quanto inverse call (or put)  could have higher or lower prices than a direct call (or put), as shown later -- see Figure \ref{fig:Payoff_Sensitivity}. Intuitively, the buyer of such a call  may seek to fix the exchange rate slightly higher or lower to her expected future exchange rate, depending on her position. Compared with their direct counterpart, a quanto inverse call loses more value as the option moves further in the money. As such, a quanto inverse call provides an affordable alternative to gain exposure to a crypto options market by paying a lower price than one would for an in-the-money (ITM) direct call of the same strike.

\begin{figure}[h!]
	\centering
	\small
	\caption{Quanto Inverse Options}
	\vspace{-0.3cm}
	\caption*{\footnotesize The payoff to a long quanto inverse call (left; blue graph) and put (right; red graph)  as a function of the settlement price. In \eqref{eq:quanto_inverse_payoff} we set $\mathbbm{Y}= \mathbbm{B}$ i.e. the notional is 1 bitcoin, and translate the inverse option payoff to USD by setting $\mathbbm{Z}=\$$. Both the quanto factor  and the option strike are set to \$25,000. }  
	\includegraphics[width=0.49\textwidth]{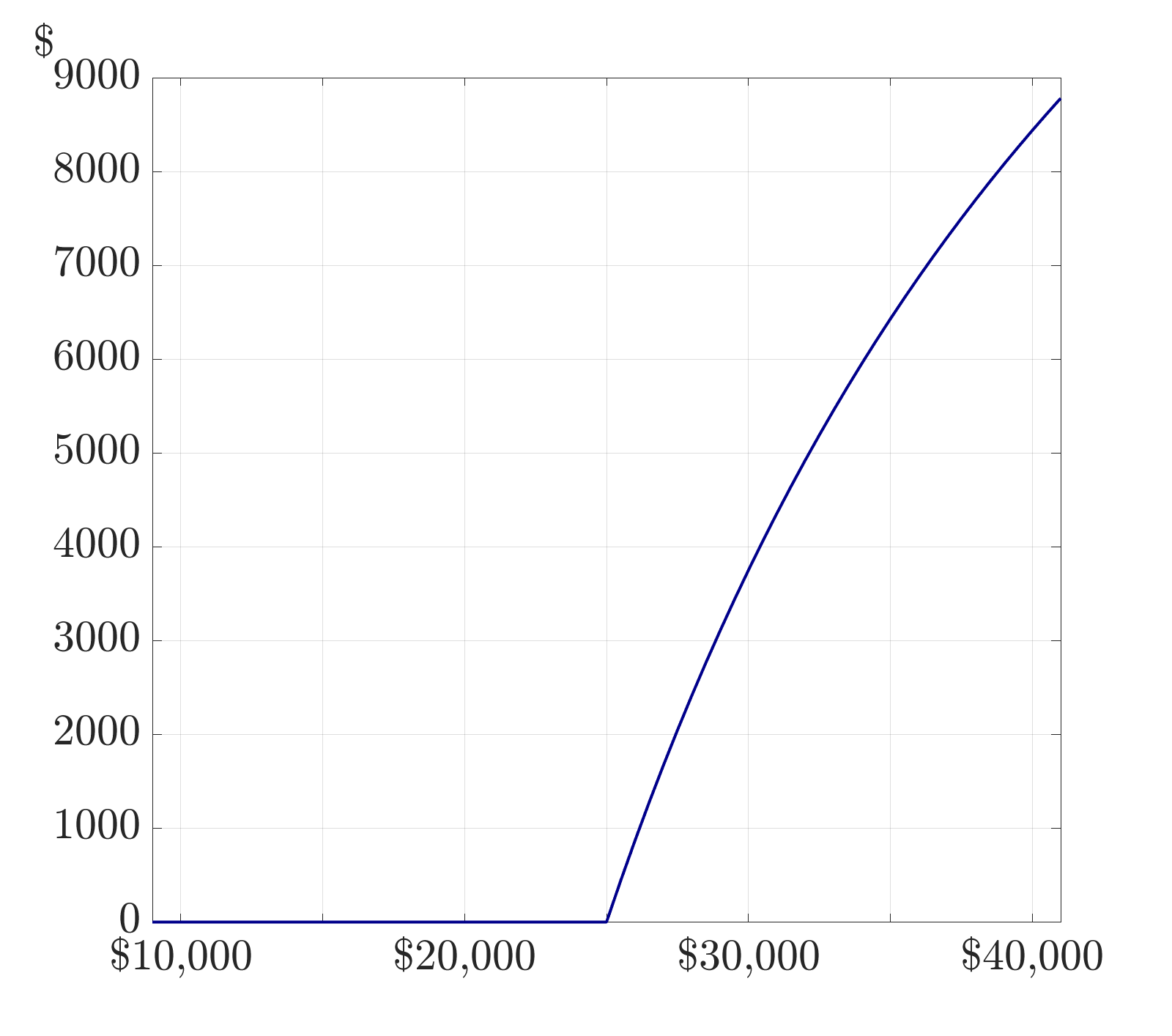}
	\includegraphics[width=0.49\textwidth]{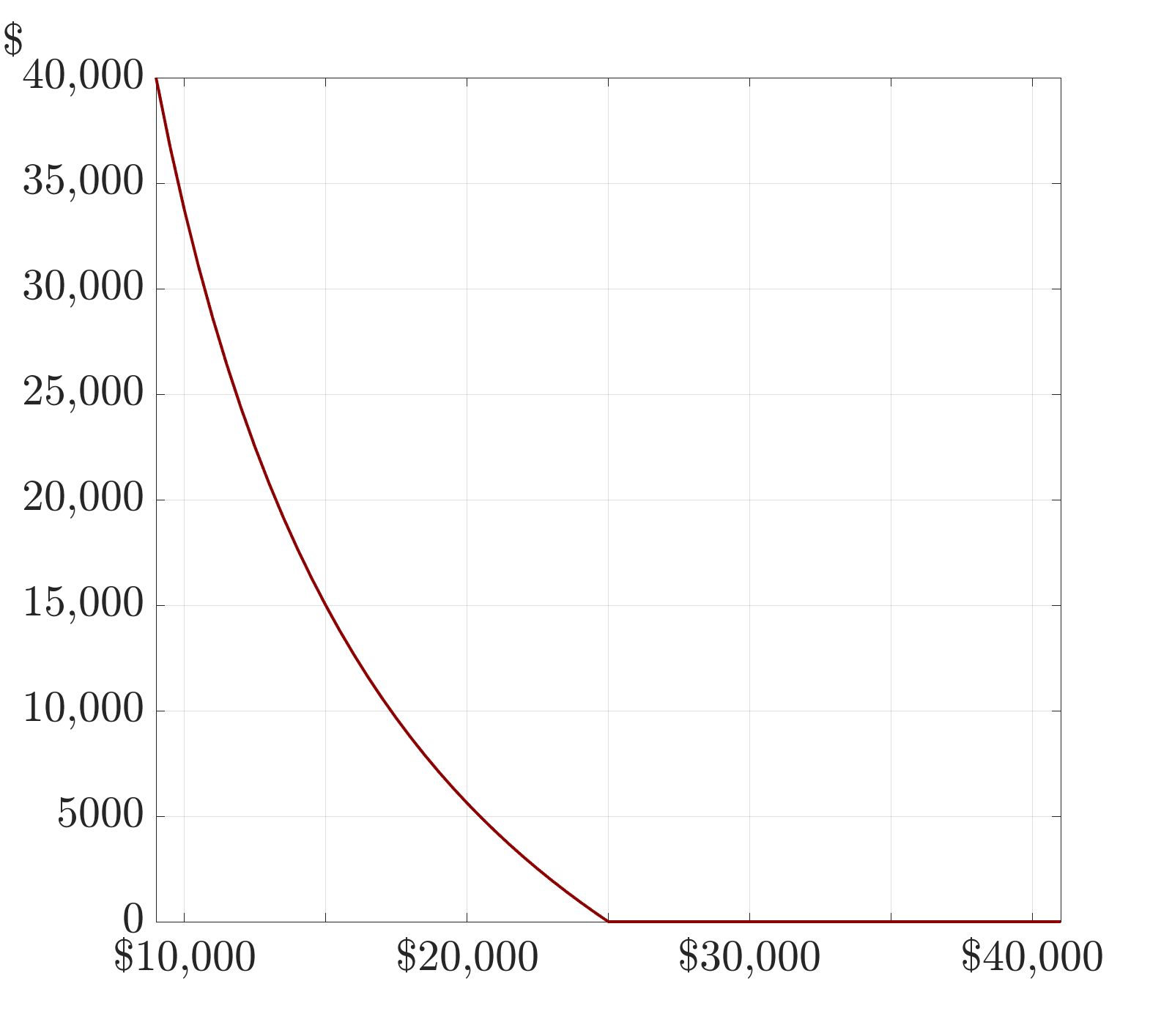}	
	\label{fig:Quanto_Inverse_Example} 
	\vspace{-13pt}
\end{figure}

Figure \ref{fig:Quanto_Inverse_Example} displays the payoffs to quanto inverse calls and puts on the BTC-USD trading pair. In this example we also use a fixed BTC/USD exchange rate (but in general the exchange-rate fix could be in BTC relative to any other crypto or fiat currency) so the shapes of these payoffs are exactly similar to the non-quanto inverse option payoff  in Figure \ref{fig:inverse_Example}. The only difference is that the vertical axis is now in USD units. In particular, the quanto inverse put payoff again increases rapidly as the underlying depreciates. This feature makes quanto inverse puts an excellent insurance against a black swan crypto event. For instance, suppose that in February 2020 a trader bought a quanto inverse put, with a notional of 1 bitcoin,  expiring on 13 March 2020. Also suppose both the strike and the quanto factor of this put were fixed at \$9,000, which is reasonable since this was about the BTC/USD rate in February 2020. On 12 March 2020 the BTC/USD rate fell to about \$3,500, so suppose this was the settlement price. Thus a standard put would have paid \$5,500 if held to expiry. But the quanto inverse put would have paid $\$9,000 \times \$5,500/3,500 = $ \$14,143, which is more than double that of the standard put.

Finally, consider how a quanto \textit{inverse} option might be constructed for a USD-denominated trader who seeks exposure to a crypto asset XYZ, but instead of trading a direct option on XYZ with currency protection against decoupling of stablecoin price from USD, as in Section \ref{sec:direct}, the trader prefers to gain exposure to an inverse option pay-off. Reasons for this could be: (1)  for the same terminal value of the underling, the quanto inverse option profit is greater than that from direct option; and (2) a quanto inverse put option payoff is not only convex but uncapped. In this case a USD-denominated trader could fix the USD price of XYZ upfront, but still retain an optional exposure to XYZ in the from of an inverse option payoff, which has payoff in XYZ.

\subsection{Summary and Comparison of Payoffs}
Table \ref{tab:payoff_comparison} summarizes the payoffs to  different types of calls and puts with and without currency protection.   As before, we use $\$$, $\mathbb{B}$ and $\mathbb{E}$ to denote payoffs in USD, bitcoin and ether, e.g. $\mathbb{B}$1 is one bitcoin or $\$$10 is ten dollars. Note that option strikes and  underlyings are always quoted in USD. The upper panel considers standard and inverse options on the exchange rate BTC/USD with strike $ K^{^{_{\$}}}=\$25,000$ and we report the payoffs for different settlement prices. The quanto versions suppose a pre-fixed  exchange rate of  $\Bar{X}^{^{_{\$/\mathbbm{B}}}}=\$22,500$. The lower panel is for options on the ETH-USD trading pair, where ETH is regarded as a security and as such could be replaced by any other crypto asset, as discussed above. In the table we let the option strike be $K^{^{_{\$}}}=\$1,750$ and again consider different scenarios for the settlement price.

\begin{table}[h!]
	\centering
	\small
	\caption{Payoffs to Standard and Inverse Options $\pm$ Currency Protection}
	\vspace{-0.3cm}
	\caption*{\footnotesize We compare the payoffs to standard and inverse calls and puts with and without the currency protection  provided by a fixed quanto factor. The currency of each payoff is recorded and numbers are rounded. The upper part is for options on BTC/USD, i.e.  with $S^{^{_{\$}}} =$ BTC/USD, with strike $ K^{^{_{\$}}}=\$25,000$ and different settlement prices from \$10,000 to \$50,000. The quanto factor for converting the standard option payoff from USD to BTC is $\Bar{X}^{^{_{\$/\mathbbm{B}}}}=\$22,500$, and for converting the inverse option payoff from BTC to USD it is $\Bar{X}^{^{_{\mathbbm{B}/\$}}}=\$22,500^{-1} =\mathbbm{B}0.0000444$. The lower part is for options on ETH-USD, i.e.  where $S^{^{_{\$}}} =$ ETH-USD,  with strike $K^{^{_{\$}}}=\$1,750$ and different settlement prices from \$500 to \$2500. The quanto factor for converting the standard option payoff from USD to BTC is again $\Bar{X}^{^{_{\$/\mathbbm{B}}}}=\$22,500$, but now for converting the inverse option payoff from ETH to USD we use $\Bar{X}^{^{_{\mathbbm{E}/\$}}}=\$2,000^{-1} =\mathbbm{E}0.0005$.}
	\begin{tabular}{l|r|ccccc}
		\toprule
		\multicolumn{1}{r}{} & \multicolumn{1}{l}{Pay-off Function} &       &       & \multicolumn{1}{c}{$S^\$_{_T}$} &       &  \\
		\cmidrule{2-7}    \multicolumn{1}{l}{Call} &       & \$10,000    & \$20,000    & \$30,000    & \$40,000    & \$50,000 \\
		\midrule
		\midrule
		$\text{Standard}$& \multicolumn{1}{c|}{$\left(S_{_T}^{^{_{\$}}} -K^{^{_{\$}}}\right)^{^{_+}}$} &   \$0    & \$0       &   \$5,000     &  \$15,000     & \$25,000  \\
		$\text{Inverse}$ & \multicolumn{1}{c|}{$\left(S_{_T}^{^{_{\$}}} -K^{^{_{\$}}}\right)^{^{_+}}/S_{_T}^{^{_{\$}}}$} &   $\mathbb{B}0$     &  $\mathbb{B}0$     &  $\mathbb{B}0.17$    &   $\mathbb{B}0.38$    & $\mathbb{B}0.5$  \\		
		$\text{Standard Quanto} $ &  \multicolumn{1}{c|}{$\Bar{X}^{^{_{\$/\mathbbm{B}}}} \left(S_{_T}^{^{_{\$}}} - K^{^{_{\$}}} \right)^{^{_+}}$} &  $\mathbb{B}0$     &  $\mathbb{B}0$     &  $\mathbb{B}0.22$    &   $\mathbb{B}0.67$    & $\mathbb{B}1.11$  \\ 
		$\text{Quanto Inverse}$ &  \multicolumn{1}{c|}{$\Bar{X}^{^{_{\mathbbm{B}/ \$}}}\left( S_{_T}^{^{_{\$}}} - K^{^{_{\$}}} \right)^{^{_+}} / S_{_T}^{^{_{\$}}}$} &   \$0    &     \$0  &   \$3,750    &    \$8,438   &  \$11,250\\
		\multicolumn{1}{r}{} & \multicolumn{1}{r}{} &       &       &       &       &  \\
		\multicolumn{1}{l}{Put} & \multicolumn{1}{r}{} &       &       &       &       &  \\
		\midrule
		\midrule
		$\text{Standard}$& \multicolumn{1}{c|}{$\left(K^{^{_{\$}}} - S_{_T}^{^{_{\$}}}\right)^{^{_+}}$}     &   \$15,000    &    \$5,000    &    \$0    &    \$0    &  \$0 \\
		$\text{Inverse}$ & \multicolumn{1}{c|}{$(K^{^{_{\$}}} - S_{_T}^{^{_{\$}}})^{^{_+}}/S_{_T}^{^{_{\$}}}$}    &    $\mathbb{B}1.5$    &   $\mathbb{B}0.25$     &   $\mathbb{B}0$     &    $\mathbb{B}0$    & $\mathbb{B}0$  \\
		$\text{Standard Quanto}  $ &    \multicolumn{1}{c|}{$\Bar{X}^{^{_{\$/\mathbbm{B}}}} \left(K^{^{_{\$}}} - S_{_T}^{^{_{\$}}} \right)^{^{_+}}$}        &    $\mathbb{B}0.67$    &    $\mathbb{B}0.22$    &   $\mathbb{B}0$      &  $\mathbb{B}0$      &  $\mathbb{B}0$ \\
		$\text{Quanto Inverse}$ & \multicolumn{1}{c|}{$\Bar{X}^{^{_{\mathbbm{B}/ \$}}} \left(K^{^{_{\$}}} - S_{_T}^{^{_{\$}}}  \right)^{^{_+}} / S_{_T}^{^{_{\$}}}$}      &   \$33,750    &   \$5,625    &   \$0    &    \$0   & \$0 \\
		\bottomrule
	\end{tabular}
	\label{tab:payoff_comparison}

\vspace{1cm}

	\begin{tabular}{l|r|ccccc}
		\toprule
		\multicolumn{1}{r}{} & \multicolumn{1}{l}{Pay-off Function} &       &       & \multicolumn{1}{c}{$S_{_T}^{^{_{\$}}}$} &       &  \\
		\cmidrule{2-7}    \multicolumn{1}{l}{Call} &       & \$500    & \$1000    & \$1500    & \$2000    & \$2500 \\
		\midrule
		\midrule
		$\text{Standard }  $ &  \multicolumn{1}{c|}{$\left(S_{_T}^{^{_{\$}}} -K^{^{_{\$}}}\right)^{^{_+}}$} &  $\$0$     &  $\$0$     &  $\$0$    &   $\$250$    & $\$750$  \\
		$\text{Inverse }  $ &  \multicolumn{1}{c|}{$\left(S_{_T}^{^{_{\$}}} -K^{^{_{\$}}}\right)^{^{_+}}/S_{_T}^{^{_{\$}}}$} &  $\mathbb{E}0$     &  $\mathbb{E}0$     &  $\mathbb{E}0$    &   $\mathbb{E}0.125$    & $\mathbb{E}0.3$  \\
		$\text{Standard Quanto} $ &  \multicolumn{1}{c|}{$\Bar{X}^{^{_{\$/\mathbbm{B}}}} \left(S_{_T}^{^{_{\$}}} - K^{^{_{\$}}} \right)^{^{_+}}$} &  $\mathbb{B}0$     &  $\mathbb{B}0$     &  $\mathbb{B}0$    &   $\mathbb{B}0.01$    & $\mathbb{B}0.03$  \\
		$\text{Quanto Inverse}  $ &  \multicolumn{1}{c|}{$\Bar{X}^{^{_{\mathbbm{E}/ \$}}} \left( S_{_T}^{^{_{\$}}} - K^{^{_{\$}}} \right)^{^{_+}} / S_{_T}^{^{_{\$}}}$} &  $\$0$     &  $\$0$     &  $\$0$    &   $\$250$    & $\$600$  \\
		\multicolumn{1}{r}{} & \multicolumn{1}{r}{} &       &       &       &       &  \\
		\multicolumn{1}{l}{Put} & \multicolumn{1}{r}{} &       &       &       &       &  \\
		\midrule
		\midrule
		$\text{Standard } $ &    \multicolumn{1}{c|}{$\left(K^{^{_{\$}}} - S_{_T}^{^{_{\$}}}\right)^{^{_+}}$}        &    $\$1,250$    &    $\$750$    &   $\$250$      &  $\$0$      &  $\$0$ \\
		$\text{Inverse } $ &    \multicolumn{1}{c|}{$\left(K^{^{_{\$}}} - S_{_T}^{^{_{\$}}}\right)^{^{_+}}/S_{_T}^{^{_{\$}}}$}        &    $\mathbb{E}2.5$    &    $\mathbb{E}0.75$    &   $\mathbb{E}0.17$      &  $\mathbb{E}0$      &  $\mathbb{E}0$ \\
		$\text{Standard Quanto} $ &    \multicolumn{1}{c|}{$\Bar{X}^{^{_{\$/\mathbbm{B}}}} \left(K^{^{_{\$}}} - S_{_T}^{^{_{\$}}} \right)^{^{_+}}$}        &    $\mathbb{B}0.06$    &    $\mathbb{B}0.03$    &   $\mathbb{B}0.01$      &  $\mathbb{B}0$      &  $\mathbb{B}0$ \\	
		$\text{Quanto Inverse}  $ &    \multicolumn{1}{c|}{$\Bar{X}^{^{_{\mathbbm{E}/ \$}}} \left( K^{^{_{\$}}} - S_{_T}^{^{_{\$}}} \right)^{^{_+}} / S_{_T}^{^{_{\$}}}$}        &    $\$5,000$    &    $\$1,500$    &   $\$333$      &  $\$0$      &  $\$0$ \\
		\bottomrule
	\end{tabular}
	\label{tab:eth_payoff_comparison}
\end{table}
The lower part of Table \ref{tab:payoff_comparison} dispays the paypff to standard and inverse calls and puts on the ETH-USD trading pair. Below this the standard quanto option payoff is for a \textit{BTC-based} trader in ETH-USD, seeking currency protection against a fall in the price of BTC relative to the dollar. The fourth payoff is for a \textit{USD-based} trader entering a quanto inverse option on ETH-USD, seeking currency protection against a fall in the price of ETH relative to the US dollar. Notice the use of notation ETH-USD means we assume ETH is security here -- and indeed could be replaced  by any other crypto  XYZ that is regarded as a security. We are particular to use notation ETH-USD in this case rather than ETH/USD, the latter specifically denoting an exchange rate in this paper. 

As already remarked, for every settlement price yielding a non-zero payoff, the standard call has a greater payoff than the quanto inverse call, but the standard put has a much smaller payoff than the quanto inverse put. This  ordering holds for both BTC/USD and ETH-USD options, as shown in the table. Furthermore, the table compares the payoffs to other types of options. For instance, in the BTC/USD case (upper panel) the inverse and standard quanto options are both denominated in BTC, and again the calls and puts have opposite ordering. That is, the inverse call has a smaller payoff than the standard quanto but the inverse put has a greater payoff than the standard quanto. Finally we note that Table  \ref{tab:payoff_comparison} only considers standard and inverse options because the direct options are similar to the standard ones, except the USD payoff is in a stablecoin.

\section{Pricing Considerations}\label{sec:pricing}
The technical issues concerning inverse option pricing focus on the debate about bitcoin -- is it a currency, commodity or a security? While the fundamental idea behind bitcoin was to facilitate a distributed payment system, and its use today reflects  a store of value to a certain extent, its high volatility suggests it acts more like a commodity or security than a currency. Other coins like ether were never created as a medium of exchange or a store of value, but rather as consumption goods.  We can picture ether as the gas for the smart contract cars that run on the highway called the Ethereum blockchain. The debate continues at the highest level of regulation and we cannot hope to bring closure to it here. Nevertheless the question whether bitcoin or ether is a security, or commodity,  or a currency is actually central to the arguments presented here. 

In the following we skip the pricing for direct and quanto options because the already voluminous extant literature still applies to these. Instead we hope to shed some light on the rather confusing properties of inverse options and present an in-depth analysis of pricing quanto inverse options. 

\subsection{Inverse Option Valuation} \label{sec:fxp} 
For this case, we consider bitcoin as a currency, and the underlying of a bitcoin option contract is a \textit{tradable} BTC/USD exchange rate,
denoted $S^\$_t$. Assuming well-functioning money markets exist for both currencies, we can define two cash bond accounts $B_t^{^{_{\$}}}=e^{r^{_{\$}}t}$ and $B_t^{^{_{\mathbb{B}}}}=e^{r^{^{_{\mathbb{B}}}}t}$ as the respective numéraires, where ${r^{^{_{\$}}}}$ and ${r^{^{_{\mathbb{B}}}}}$ are the risk-free interest rates in the corresponding currencies. Furthermore, let $\left(\Omega, \mathcal{F},\mathbb{P} \right)$ be a filtered probability space with filtration $\mathcal{F}$ and probability measure $\mathbb{P}$. We assume log-normal asset dynamics under the physical measure $\mathbb{P}$:
\begin{equation}\label{eq:FX_USD_drift}
\frac{dS^{^{_{\$}}}_t}{S^{^{_{\$}}}_t} = \mu^{^{_{\$}}}dt+\sigma^{^{_{\$}}}dW_t^{^{_{\$}}},
\end{equation}
where $\mu^{^{_{\$}}}$ is the drift, $\sigma^{^{_{\$}}}$ is the volatility, and $W^{^{_{\$}}}$ denotes a standard Brownian motion. In this framework, we can rewrite the payoff  $V_{_T}^{^{_{\mathbb{B}}}}$ to a $T$-maturity  vanilla call on $S_{_T}^{^{_{\$}}}$ from (\ref{eq:inverse_payoff}) to:
\begin{equation}\label{eq:FX_Payoff}
V_{_T}^{^{_{\mathbb{B}}}}=  K^{^{_{\$}}} \left( K^{^{_{\mathbb{B}}}} - S_{_T}^{^{_{\mathbb{B}}}} \right)^+,
\end{equation}
where $S^{^{_{\mathbb{B}}}}= \frac{1}{S^{^{_{\$}}}}$ the opposite exchange rate, i.e. USD/BTC. Note that a USD-denominated call becomes a BTC-based put with strike $K^{^{_{\mathbb{B}}}}$ being the inverse of the USD-strike price. Note that the option can be priced trivially via standard FX option pricing formulae, and satisfies the put-call duality conditions as articulated, for example, in \cite{GK1983}. The price of the put in BTC is given by:
\begin{align} \label{eq:FX_Put_Price}
	P_t^{^{_{\mathbb{B}}}} &= e^{-r^{^{_{\mathbb{B}}}} \tau} K^{^{_{\mathbb{B}}}}  \Phi \left(-d_2^{^{_{\mathbb{B}}}} \right)  - e^{-r^{^{_{\$}}} \tau } S^{^{_{\mathbb{B}}}}_t  \Phi \left(-d_1^{^{_{\mathbb{B}}}}  \right),  \\
	d_1^{^{_{\mathbb{B}}}} &= \frac{1}{\sigma^{^{_{\mathbb{B}}}} \sqrt{\tau}} \left[ \ln \left(\frac{S_t^{^{_{\mathbb{B}}}} }{K^{^{_{\mathbb{B}}}}} \right)+ \left(r^{^{_{\mathbb{B}}}} - r^{^{_{\$}}} + \frac{(\sigma^{^{_{\mathbb{B}}}})^2}{2}\right) \tau \right] , \nonumber \\
	d_2^{^{_{\mathbb{B}}}} &= d_1^{^{_{\mathbb{B}}}} - \sigma^{^{_{\mathbb{B}}}}\sqrt{\tau}. \nonumber 
\end{align}
Similarly, we can express the variables in a USD-denominated framework:
\begin{align*}
	d_1^{^{_{\mathbb{B}}}} &= \frac{1}{\sigma^{^{_{\$}}} \sqrt{\tau}} \left[ -\ln \left(\frac{S_t^{^{_{\$}}} }{K^{^{_{\$}}}} \right)- \left(r^{^{_{\$}}} - r^{^{_{\mathbb{B}}}} - \frac{(\sigma^{^{_{\$}}})^2}{2}\right) \tau \right] =-d_2^{^{_{\$}}}, \\
	d_2^{^{_{\mathbb{B}}}} &= -d_1^{^{_{\$}}}.
\end{align*}
Clearly, $K^{^{_{\$}}}$ units of BTC-denominated puts are equivalent to $\frac{1}{S_t^{^{_{\$}}}}$ units of USD-denominated calls:
\begin{align} \label{eq:pcd}
 K^{^{_{\$}}} P_t^\mathbb{B} = e^{-r^{^{_{\mathbb{B}}}} \tau }  \Phi \left(-d_2^{^{_{\mathbb{B}}}}  \right)  - e^{-r^{^{_{\$}}} \tau} K^{^{_{\$}}} S^{^{_{\mathbb{B}}}}_t  \Phi \left(-d_1^{^{_{\mathbb{B}}}}  \right) = \frac{1}{S^{^{_{\$}}}} C_t^\mathbb{\$}.
 \end{align}
{This pricing approach can be readily extended beyond Black-Scholes world. Solutions to option pricing problems are attainable, at least in the Fourier transform sense, for any tractable L\'evy processes governing the evolution of the USD/BTC exchange rate. Popular choices are the Heston model, the SABR model and their multi-factor and/or jump-extended variants, among others. Recently, \cite{TH2022} calibrated a correlated jump extended Heston model to Deribit BTC inverse options.}

Crucially, this simple pricing approach relies on the assumptions that both markets denominated in USD and in BTC are complete, and there are no restrictions in exchanging wealth from one to the other. However, considering the current state of bitcoin options, we must note that this assumption is not satisfied. First, not only is there no well-functioning money market for bitcoin, it is non-existing at all.\footnote{Risk-free lending platforms are the closest which come to some kind of money market. These companies promised risk-free interest if a client is willing to deposit his/her crypto. However, poor risk management and a dangerous mixture of incompetence and ignorance brought major lenders to their knees just recently. \href{https://www.ft.com/content/8d6dee7d-2cc9-4663-a0a2-e469686baca5}{Celsius} and \href{https://www.ft.com/content/0b5b68d9-85f1-47ce-a9f7-34252e4fe2ce}{Voyager} are the latest big players filing for bankruptcy shattering trader's trust in the whole crypto ecosystem. } Second, the exchanges on which inverse options are traded (which accept no fiat and are unregulated) use non-traded underlyings. For example, the Deribit bitcoin inverse options use their own `Deribit Bitcoin Index' as the underlying, which is an average of bitcoin spot prices from (currently) eleven  different  centralised exchanges. Due to its frequent rebalancing, the physical replication of this index is an immensely difficult and expensive task. An agent would be required to hold bitcoin positions on multiple exchanges and rebalance these constantly. Moreover, the final option settlement price is the average of the index during the last 30 minutes before expiry. This important feature about Deribit inverse options is often ignored. But the price of bitcoin can change considerably during 30 minutes -- much more than we see for traditional financial instruments. Hence, the underlying is not tradable and the market is incomplete. Therefore, equation (\ref{eq:FX_Put_Price}) provides only an approximate option price, even in a Black-Scholes world.

To fully understand the valuation and hedging of bitcoin options on Deribit, we need to consider the actual state of the cryptocurrency option market. To circumvent the difficulties in reconstructing and trading the underlying index of the Deribit bitcoin options we consider instead the Deribit perpetual futures, because this tracks the index closely through the funding mechanism. However, this introduces an unhedgeable basis risk that can be sizable in a volatile market. Now, option pricing in an incomplete market in the presence of basis risk is typically solved through indifference pricing by formulating a stochastic control problem in the mean of the Hamilton-Jacobi-Bellman (HJB) partial differential equation  \citep{Monoyios2004245}, or by solving the corresponding forward-backward stochastic differential equation \citep{Rouge2000259}. Consider the probability space $\left(\Omega, \mathcal{F},\mathbb{P} \right)$ as in the standard inverse pricing case. The non-tradable underlying asset (in this case the Deribit bitcoin index, $S^{^{_{\mathbb{B}}}}_t$) and the hedging instrument (the perpetual futures, $Y^{^{_{\mathbb{B}}}}_t$) evolve according to the geometric Brownian motions:
\begin{align}\label{eq:Incomplete_BTC_Dynamics}
	\frac{dS^{^{_{\mathbb{B}}}}_t}{S^{^{_{\mathbb{B}}}}_t}& = \mu_{_{\mathbb{B}}}dt + \sigma_{_{\mathbb{B}}}dW_t^{^{_{\mathbb{B}}}}, \\
		\frac{dY^{^{_{\mathbb{B}}}}_t}{Y^{^{_{\mathbb{B}}}}_t}& = \mu dt + \sigma dW_t,
\end{align}
  where  $W^{^{_{\mathbb{B}}}}$ and $W$ correlated  standard Brownian motions with  $ \langle dW^{^{_{\mathbb{B}}}}, dW \rangle = \rho\, dt$.  Note that $S$ and $Y$ are not perfectly correlated, i.e. $\mid \rho \mid <1$ and hence we cannot find a unique equivalent martingale measure (EMM) for which the discounted value of $Y$ is a martingale under risk-neutral measure. We rewrite
  $$W^{^{_{\mathbb{B}}}}_t= \rho W_t + \epsilon 	\hspace{-0.27em}\tildea{W}_t^{_{_{^{_{\mathbb{B}}}}}}, $$ 
  where $\epsilon = \sqrt{1-\rho^2}$ with independent $	\tildea{W}_t^{^{_{\mathbb{B}}}}$ and $W_t$. We further denote by $\left\{\mathcal{G}_t\right\}_{0 \leq t \leq T}$ the filtration generated by $W_t^{^{_{\mathbb{B}}}}$. Note that this Brownian motion drives the non-traded inverse index asset.

Assume there exists an equivalent measure $\mathbb{Q}$ to $\mathbb{P}$ on $\mathcal{F}$. Then there exists adapted processes $m_{_T}$ and $g_{_T}$ where the Radon-Nykodym derivative is given by:
\begin{equation}\label{eq:Incomplete_Radon_Nykodym}
	\frac{d\mathbb{Q}}{d \mathbb{P}}= M_t, 
\end{equation}
where $\left\{M_t\right\}_{0 \leq t \leq T}$ is a  $\mathbb{P}$-martingale with  $M_t$ is given by:
\begin{equation}\label{eq:Incomlete_M}
	M_t = \exp \left[ \int_{0}^{t}  m_u dW_u + \int_{0}^{t} g_u	d \hspace{-0.27em}\tildea{W}^{_{_{^{_{\mathbb{B}}}}}}_u -  \frac{1}{2} \int_{0}^{t} m_u^2 du - \frac{1}{2} \int_{0}^{t} g^2_u du \right].
\end{equation}
Using the multidimensional Girsanov theorem we see that the processes $\left\{\tildea{W}_t,\hata{W}^{_{_{_{^{_{\mathbb{B}}}}}}}_t \right\}_{0 \leq t \leq T}$ defined by:
\begin{align}
	\begin{pmatrix}
		\tildea{W}_t \\
	\hata{W}^{_{_{_{^{_{\mathbb{B}}}}}}}_t 
	\end{pmatrix} 
	= 
	\begin{pmatrix}
		W_t + \int_{0}^{t} m_u du \\
	\tildea{W}_t^{_{_{^{_{\mathbb{B}}}}}} + \int_{0}^{t } g_u du
	\end{pmatrix} ,
\end{align}
is an independent Brownian motion under $\mathcal{Q}$.  Further, the dynamics under  $\mathcal{Q}$  are given by: 
\begin{align}\label{eq:Incomplete_BTC_Dynamics}
	\frac{dS^{^{_{\mathbb{B}}}}_t}{S^{^{_{\mathbb{B}}}}_t}& = \left( \mu_{\mathbb{B}} - \sigma_{\mathbb{B}} \left(\rho m_t +\epsilon g_t \right) \right) dt +  \sigma_{\mathbb{B}} dW_t, \\
	\frac{dY^{^{_{\mathbb{B}}}}_t}{Y^{^{_{\mathbb{B}}}}_t}& =  \left( \mu - \sigma m_t \right) dt +  \sigma d \hspace{-0.27em} \hata{W}_t,
\end{align}
where $\hata{W}_t = \rho \hspace{-0.27em}\tildea{W}_t + \epsilon \hspace{-0.27em} \hata{W}^{_{_{_{^{_{\mathbb{B}}}}}}}_t$ is a Brownian motion such that $ \langle d \hspace{-0.27em}\tildea{W}_t, d \hspace{-0.27em} \hata{W}_t\rangle = \rho \,dt$. For $\mathbb{Q}$ be be a local EMM, $Y_t$ needs to be a $\mathbb{Q}$-local martingale, i.e. iff:
$$ \mu - m_t \sigma = r \quad \Rightarrow \quad m_t = m:= \frac{\mu -r}{\sigma}. $$
Note that the EMM is uniquely defined. On the other hand $X$ is non-tradable which lets $g_{_T}$  be of any arbitrary form which results in an infinite set of possible EMM. We define this set as $\mathcal{M}$ which is in correspondence with the set of $g_t$. We further want to link the local martingale with an equivalent probability measure. Denote an equivalent measure  $ \tildea{\mathbb{P}}$ to  $\mathbb{P}$ on $\mathcal{G}$, the risk-neutral density given by:
\begin{align}
	\frac{d\hspace{-0.27em}\tildea{\mathbb{P}}}{d\mathbb{P}} &= \tildea{m}_t, \\
	\tildea{M}_t &= \exp\left\{   -\int_{0}^{t} \theta_u dW^{^{_{\mathbb{B}}}}_u -\frac{1}{2} \int_{0}^{t} \theta^2_u du \right\},
\end{align}
where $\theta_t$ is a $\mathcal{G}_t$-adapted process. Under $\tildea{\mathbb{P}}$ we define $\hata{W}$ as:
\begin{align}
	\hata{W}_t &= W^{^{_{\mathbb{B}}}}_t+ \int_{0}^{t} \theta_u du
\end{align}
and underlying dynamics under $\tildea{\mathbb{P}}$ given by:
\begin{align}
	\frac{dS^{^{_{\mathbb{B}}}}_t}{S^{^{_{\mathbb{B}}}}_t} &= \left( \mu_{_{\mathbb{B}}} -\sigma_{_{\mathbb{B}}} \theta_t \right)dt +\sigma_{_{\mathbb{B}}} d \hspace{-0.27em} \hata{W}_t.
\end{align}
Note that the dynamics of the non-tradable inverse index are the same under $\mathbb{Q}$ and $\tildea{\mathbb{P}}$ when the integrands $m_t, g_t$ and $\theta_t$ are given as: $$ \rho m_t + \epsilon g_t = \theta_t.$$
 
 Many arbitrage valuation models such as the BSM-model rely on the assumption that any option's claim can be replicated by a portfolio and hence hedged perfectly. The absence of such a hedging portfolio leads to major difficulties when evaluating options in this fashion. A preference-dependent approach is therefore necessary for pricing these options. By making assumptions about the shape of the trader's utility function  and their initial wealth endowments, the option pricing problem can be transformed into an optimal trading strategy problem where the trader seeks to maximise her expected utility with and without trading in options contracts. Many papers already address this topic in much detail -- see \cite{Rouge2000259}, \cite{Monoyios2004245}, \cite{AIR2010} and \cite{Davis2000} and many others. This way it can be presented as the numerical solution to the corresponding HJB partial, or the forward-backward stochastic, differential equation. However, there are significant limitations to using this type of model for pricing and hedging purposes. The option pricing formula and the associated Greeks depend on whether the option position is long or short, and the type of utility functions used can often be restrictive -- in some cases it would even yield option prices that depend on the agent's initial wealth.

\subsection{Quanto Inverse Option Valuation}\label{sec:recip} 
 It's worth noting upfront that, the quanto inverse option tabulated in Table (\ref{tab:payoff_comparison}) shares a  payoff structure similar to the inverse option payoff, but  is denominated differently. This subtle difference has important consequences in terms of how these options should be priced. Consider first a standard FX call on /USD that matures at time $T$, with either a USD-denominated payoff:
\begin{equation}\label{eq:FX_Call_USD_Payoff}
	V^{^{_{\$}}}_{_T}= \left(S^{^{_{\$}}}_{_T} - K^{^{_{\$}}}\right)^+,
\end{equation}
or a BTC-denominated payoff:
\begin{align} \label{eq:FX_Call_BTC_Payoff}
V_{_T}^{^{_{\mathbb{B}}}}=  K^{^{_{\$}}} \left( K^{^{_{\mathbb{B}}}} - S_{_T}^{^{_{\mathbb{B}}}} \right)^+.
\end{align}
We have already shown that this option can be priced in a standard FX option pricing framework. Now consider an exotic option that pays:
\begin{align} \label{eq:Reci_put_USD_Payoff}
V_{_T}^{\mathbb{\$}}=  \Bar{X}^{^{_{\mathbbm{B}/\$}}}\frac{\left(S^{^{_{\$}}}_{_T} - K^{^{_{\$}}}\right)^+}{S^{^{_{\$}}}_{_T}},
\end{align}
where $\Bar{X}^{^{_{\mathbbm{B}/\$}}}$ is a predetermined exchange rate that transforms the bitcoin denominated point value of an standard inverse option to a USD payoff. In the following we simplify notation and omit the superscript, so we set   $\Bar{X}^{^{_{\mathbbm{B}/\$}}} = \Bar{X}$. Similar to  CME options, these would be USD-settled options, on the exchange rate of any token, or a reference rate or a futures contract. In this example, we focus on the USD value of one bitcoin. Using the same notation as in Section \ref{sec:quantoinverse} we denote the underlying of this product at maturity by  $S^{^{_{\$}}}_{_T}$, and highlight its USD-denomination via the superscript. Without losing generality, we assume the dividend yield is zero, and the US money market provides a risk-free interest rate $r$. The risk-neutral $\mathbb{Q}$-dynamics of the underlying are given by:
\begin{align*}
	\frac{dS_t^{^{_{\$}}}}{{S_t^{^{_{\$}}}}}&= r dt+ \sigma dW_t,\\
	S^{^{_{\$}}}_{_T} &= S^{^{_{\$}}}_t \exp\left\{\left(r-\frac{1}{2}\sigma^2\right)(T-t) + \sigma (W_T-W_t) \right\}.
\end{align*}
Now let  $Y_t^{^{_{\$}}} =\left(S_t^{^{_{\$}}}\right)^{{_{-1}}}$. By It\^{o}'s lemma the $\mathbb{Q}$-dynamics of $Y_t^{^{_{\$}}}$ are:
\begin{equation}
	\frac{dY_t^{^{_{\$}}}}{Y_t^{^{_{\$}}}}= \left(\sigma^2-r\right) dt- \sigma dW_t.
\end{equation}
The value of the discounted price process is therefore: 
$$\tildea{Y}_t =  e^{-r\tau} \EX^{\mathbb{Q}}\left[Y_{_T}^{^{_{\$}}}\mid \mathcal{F}_{_T}\right]  = Y_{_T}^{^{_{\$}}} e^{\left(\sigma^2 -2r\right)\tau}$$
where  $\tau = T - t$. We note that $\tildea{Y}_t$ is a martingale under the $\mathbb{Q}$-measure and hence we perform a change of numéraire from the risk-neutral measure $\mathbb{Q}$ to an equivalent martingale measure $\tildea{\mathbb{Q}}$ where $\tildea{Y}_t$ is the new numéraire. We can rewrite the dynamics of $S_t^{^{_{\$}}}$ as:
\begin{align*}
	\frac{dS_t^{^{_{\$}}}}{S_t^{^{_{\$}}}} &= \left(r-\sigma^2\right)dt+ \sigma d\hspace{-0.27em}\tildea{W}_t,\\
	S^{^{_{\$}}}_t&=S^{^{_{\$}}}_0 \exp\left\{\left(r-\frac{3}{2}\sigma^2\right)t +\sigma \hspace{-0.27em}\tildea{W}_t\right\},             
\end{align*}
where $\tildea{W}_{_T} = {W_t} + \sigma t $ is a Wiener process under the martingale measure $\tildea{\mathbb{Q}}$. Under these assumptions, we can express the Radon-Nikodym derivative by:
\begin{equation*}
	\left.\frac{d \hspace{-0.27em}\tildea{\mathbb{Q}}}{d\mathbb{Q}}\right\rvert_{t} = 
	\exp\left\{\frac{1}{2}\sigma^2t+ \sigma W_t\right\}.
\end{equation*}

Denoting the price of the quanto inverse call at $t$ by $C_t^{\mathrm{q}}$, its valuation under the new measure is trivial since it is a plain vanilla option and follows the exact same steps as the risk-neutral derivation of the Black-Scholes formula: 
\begin{align} 
C^{\mathrm{q}}_t &=  \frac{B_t}{B_{_T}}	\EX^{{\mathbb{Q}} }_t\left[ \Bar{X}\left(\frac{{S^{^{_{\$}}}_{_T}}-K^{^{_{\$}}}}{{S^{^{_{\$}}}_{_T}}}\right)^+\right] = \EX^{\tildea{\mathbb{Q}} }_t\left[ \left(\left.\frac{d \hspace{-0.27em} \tildea{\mathbb{Q}}}{d\mathbb{Q}}\right\rvert_{_T}\right)^{_{-1}}  \frac{ \Bar{X}}{{S^{^{_{\$}}}_{_T}}} \left({S^{^{_{\$}}}_{_T}}-K^{^{_{\$}}}\right)^+ \right] \nonumber \\
	&= e^{\left(\sigma^2 -2r\right)\tau}\frac{ \Bar{X}}{{S^{^{_{\$}}}_t}} \EX^{\tilde{\mathbb{Q}} }_t \left[\left({S^{^{_{\$}}}_{_T}} - K^{^{_{\$}}}\right)^+\right] = e^{\left(\sigma^2 -2r\right)\tau}\frac{ \Bar{X}}{{S^{^{_{\$}}}_t}} \int_{-\infty}^{\infty} \left( S^{^{_{\$}}}_{_T}(z) - K^{^{_{\$}}}\right)^+ \phi(z)dz, 
	\label{eq:Inverse_Call_Pricing}
\end{align}	
where $ {S}^\$_{_T}(z) = {S}^\$_t \exp \left\{ \frac{1}{2}\sigma^2 \tau +\sigma \sqrt{\tau} z\right\}$,  $z$ is drawn from a standard normal distribution, and $\phi(\cdot)$ is the corresponding probability density function. Note that $$\left(S^{^{_{\$}}}_{_T}(z)-K^{^{_{\$}}}\right)^+ = 0 \Leftrightarrow z\geq \frac{\ln \left(\frac{S^{^{_{\$}}}_t}{K^{^{_{\$}}}}\right)+\left( r- \frac{3}{2}\sigma^2\right)\tau}{\sigma \sqrt{\tau}}=d_3.$$
Thus, using $\Phi(\cdot)$ to denote the standard normal distribution function we can calculate the integral on the RHS of (\ref{eq:Inverse_Call_Pricing}):
\begin{align*}
 \int_{-\infty}^{d_3}\left(S^{^{_{\$}}}_{_T}(z)-K^{^{_{\$}}} \right) \phi(z)dz = S^{^{_{\$}}}_t \int_{-\infty}^{d_3} \exp \left\{ \left( r- \frac{3}{2}\sigma^2\right) \tau +\sigma \sqrt{\tau} z\right\}   \phi(z)dz - K^{^{_{\$}}} \Phi(d_3).
\end{align*}
Evaluating the integral yields the GBM price of the inverse call with strike $K^{^{_{\$}}}$ as:
\begin{equation}
C^{\mathrm{q}}_t= e^{-r\tau} \Bar{X} \left[ \Phi (d_2) - e^{\left(\sigma^2-r\right) \tau } Y^{^{_{\$}}}_t K^{^{_{\$}}} \Phi(d_3)\right], 
\end{equation} \label{eq:call}
where $d_2 = d_3 +\sigma \sqrt{\tau}$ is the same as we see in the Black-Scholes FX pricing formula, i.e. $d_2 = d_2^{_{\$}}$ under the assumption $r^{_{\mathbb{B}}}=0$. A similar argument yields the time $t$ GBM price of an inverse put with strike $K^{^{_{\$}}}$ USD and maturity $T$ as:
\begin{equation}
P^{\mathrm{q}}_t= e^{-r\tau} \Bar{X} \left[ e^{\left(\sigma^2 -r\right)\tau }  Y^{^{_{\$}}}_t K^{^{_{\$}}}\Phi(-d_3) -\Phi(-d_2)\right]. 
	\end{equation} \label{eq:put}
	
The two likelihood functions $\frac{d\mathbb{Q}^{^{_{\mathbb{B}}}}}{d\mathbb{Q}^{^{_{\$}}}}$ and $	\frac{d \tildea{\mathbb{Q}}}{d\mathbb{Q}}$, albeit  very similar, result in different pricing and hedging properties for the two functions, driven by the difference in  denomination between these two contracts. In the quanto inverse option case it is possible, and indeed more convenient, to start from the risk-neutral dynamics of the underlying because the payoff is denominated in USD. One may define the equivalent martingale measure for the quanto inverse of the underlying denominated in USD, and price the option accordingly. However, this approach is inappropriate for an inverse option when the payoff is denominated in BTC. Due to Siegel's exchange paradox, denomination conversion from USD to BTC should be performed first under $\mathbb{P}$. Then a risk-neutral measure in BTC can  be established which is symmetrical to the risk-neutral measure in USD. This would be the appropriate measure for pricing derivatives denominated in BTC.  

\begin{figure}[h!]
	\centering
	\caption{Inverse and Quanto Inverse Option Prices under GBM}
	\caption*{\footnotesize Option payoffs and prices obtained using the Black-Scholes formula for the inverse options and our formula \eqref{eq:Inverse_Call_Pricing} for the quanto inverse options. Prices are represented as a function of the underlying price with a thicker line as the option approaches expiry. Time to maturities of 10 days, 3 months, 6 months and 1 year are shown. In these plots the first column displays calls in blue and the second displays puts in red;  the upper row shows the inverse option prices and payoffs, and the lower shows the quanto inverse option prices and payoffs.  All four plots are calculated using the same $K=\$25,000$, $r=0\%$ and $\sigma = 75\%$ for all maturities but with different USD-denominated contingent claims (\ref{eq:inverse_payoff}) and (\ref{eq:quanto_inverse_payoff}), respectively. For the quantos we set $\Bar{X}=\$22,500$.} 
	\includegraphics[width=\textwidth]{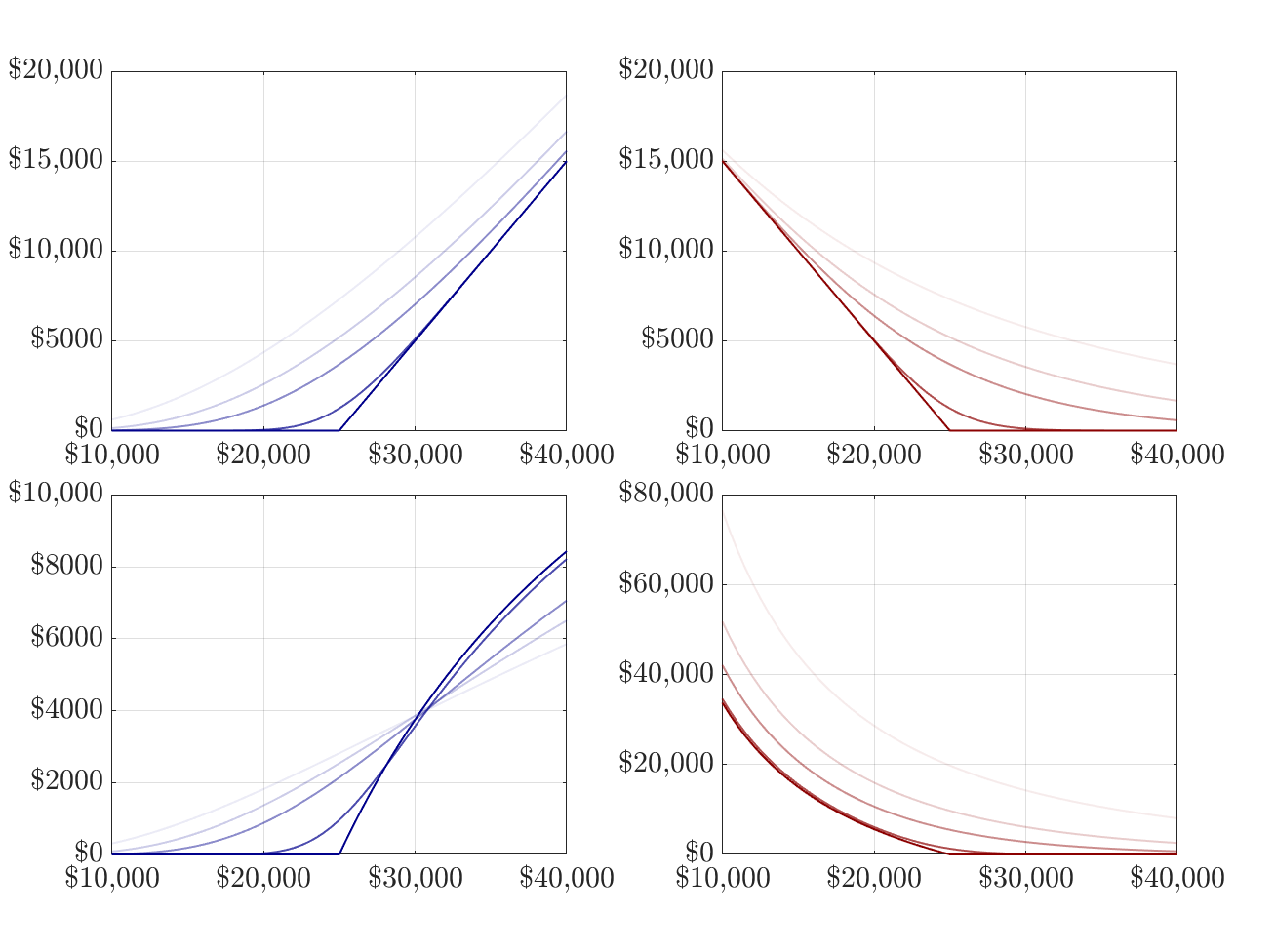}
	\label{fig:payoffcomp}
\end{figure} 

Figure \ref{fig:payoffcomp} compares (a) the USD-denominated inverse BSM prices and payoffs (above) with (b) the prices and payoffs for quanto inverse options (below). These are displayed above as a function of ${S}^\$_{_T}$, and we display prices for different maturities ranging from ten days to a year. The panels on the left show calls (blue) and on the right we have the puts (red). The inverse option payoffs have the familiar convex structure of the BSM pricing function, with an increasing gamma and a positive vega, and the price approaches the payoff as the option approaches expiry. The longer the time to expiration, the more valuable the options, i.e. the theta is positive.

The quanto inverse option pricing functions depicted in the lower panels of Figure \ref{fig:payoffcomp} behave very differently. The payoff to a quanto inverse call is capped above at $\Bar{X}$. Deep ITM quanto inverse calls decrease in value as the time to maturity increases, where they are valued below their intrinsic values. Indeed, very deep ITM quanto inverse call prices could be much lower than one would think by simply looking at ATM option prices.\footnote{The area around ATM and slightly ITM quanto inverse calls is extremely sensitive. At maturity, if the underlying price is 10\% higher than the strike, the payoff would be roughly \$2050; if the underlying price is twice the strike the payoff will be \$11,250; and if the underlying price was 10 times the strike, the payoff would be \$20,250.}  Furthermore, the convexity of the quanto inverse call pricing function changes as the underlying price increases: it starts as a convex function but then changes into a concave function as the option moneyness increases. Therefore the delta of a quanto inverse call is not a monotonic increasing function with respect to the underlying price, as it is for vanilla options. There exists a global maximum at which gamma changes from being positive to being negative. For deep OTM and ATM options, the term structure of quanto inverse calls resembles that of inverse calls but  for deep ITM options this pivots. Counter-intuitively, the price of a short-term deep ITM option exceeds the prices of their long-term counterparts, indicating a \textit{negative} theta for these moneyness levels. 

The payoff to the quanto inverse put in Figure \ref{fig:payoffcomp} is uncapped with the  put price tending towards $\infty$ as the underlying price falls, whereas the payoff to a non-quanto inverse put converges to $K$. For all moneyness  and maturities, the put pricing curves are convex decreasing with the strongest price sensitivity for ATM options.  Unlike the inverse put delta, the quanto inverse put delta  decreases monotonically with the underlying price, but it has no lower bound.  The theta  for a quanto inverse put is positive, like its inverse option counterpart, i.e.  the longer the time to maturity, the more valuable the option. It is interesting that the roles of calls and puts are now reversed, in that an inverse call and a quanto inverse put can -- theoretically -- pay an infinite amount to the holder, whereas the inverse put and quanto inverse call are capped at $K$ and $\Bar{X}$, respectively.

\begin{figure}[h!]
	\centering
	\small
	\caption{Payoff Sensitivity with Respect to $\Bar{X}^{^{_{\mathbbm{B}/\$}}}$}
	\vspace{-0.3cm}
	\caption*{\footnotesize Payoff comparison of inverse and quanto inverse call (left column) and put (right column) with different predetermined exchange factor $\Bar{X}$. The underlying at maturity ranges between $\$10,000$ and $\$40,000$ with a strike price at $K^{^{_{\$}}}$ = \$25,000. We distinguish between inverse (black) and quanto inverse (red) options and compare different $\Bar{X}$ between each column, i.e.  $\Bar{X}=\$15,000$ (upper row) and $\Bar{X}=\$35,000$ (lower row). }  
	\includegraphics[width=0.9\textwidth]{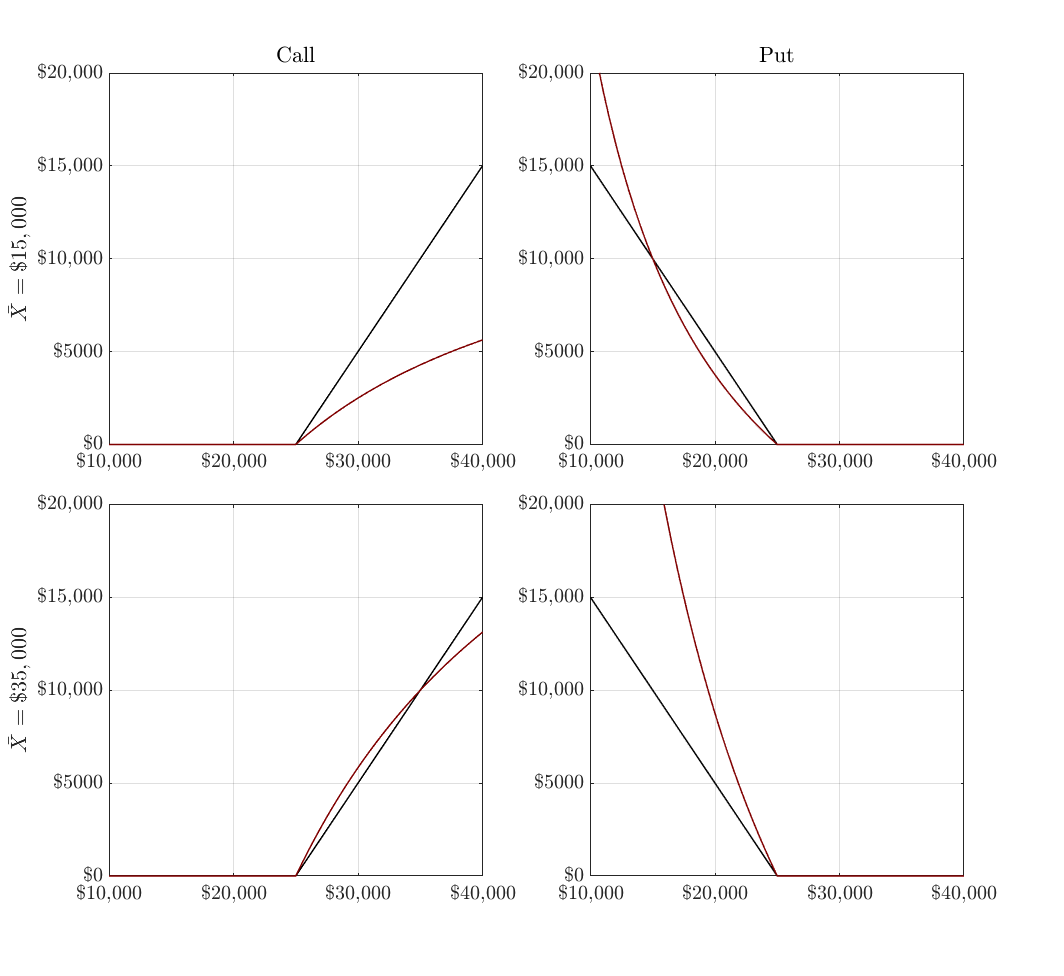}
	\label{fig:Payoff_Sensitivity}
	\vspace{-26pt}
\end{figure}

Now we investigate the quanto inverse option price  dependence on  the prefixed conversion factor $\Bar{X}$. Figure (\ref{fig:Payoff_Sensitivity}) compares  the maturity payoffs to inverse and quanto inverse options for two different values of $\Bar{X}$. Suppose an agent who is optimistic, but not euphoric, about future returns enters a long position in a quanto inverse call. She has the greatest interest in negotiating a quanto factor as high as possible because this increases the slope of the terminal payoff. For instance, the bottom left graph in Figure (\ref{fig:Payoff_Sensitivity}) illustrates how the quanto inverse call payoff can exceed that of the inverse call when  $\Bar{X}$ is high. The advantage of this long position depends on the underlying having a maturity value between $K$ and $\Bar{X}$: if the underlying ends below the strike at maturity the option expires worthless; and above $\Bar{X}$ the agent would profit more by entering a direct (or standard) option. Market makers could earn a premium by offering these two alternatives to traders seeking to profit from high future volatility. Now consider the inverse vs quanto inverse put, on the right in Figure \ref{fig:Payoff_Sensitivity}. Here it is the writer not the buyer of the option who can use  quanto inverses to their advantage. Assuming the quanto inverse put ends up ITM with $\Bar{X}<<K$ (top right plot), then the sell side would reduce their losses up to $\Bar{X}$ but exponentially increase them afterwards. We conclude that the premium on the quanto inverse options could be relatively low to write a call, but would need to be exceptionally high to write a put because, in theory, there is a non-zero probability of the asset price  reaching zero resulting in an infinite loss for the writer of quanto inverse puts.

\begin{figure}[h!]
	\centering
	\small
	\caption{Volatility-Maturity Price Sensitivity}
	\vspace{-0.3cm}
	\caption*{\footnotesize Comparison of call prices as a function of the underlying using the inverse pricing formula (blue) and the quanto inverse pricing formula (red), given in USD. The figures depict different times to maturity (ranging from ten days to six months) under different constant volatilites (ranging from 50\% to 200\%). We assume constant zero interest rate and fix both the underlying and $\Bar{X}$ at \$25,000. We display only the 20\% area around the strike price.  }
	\includegraphics[width=\textwidth]{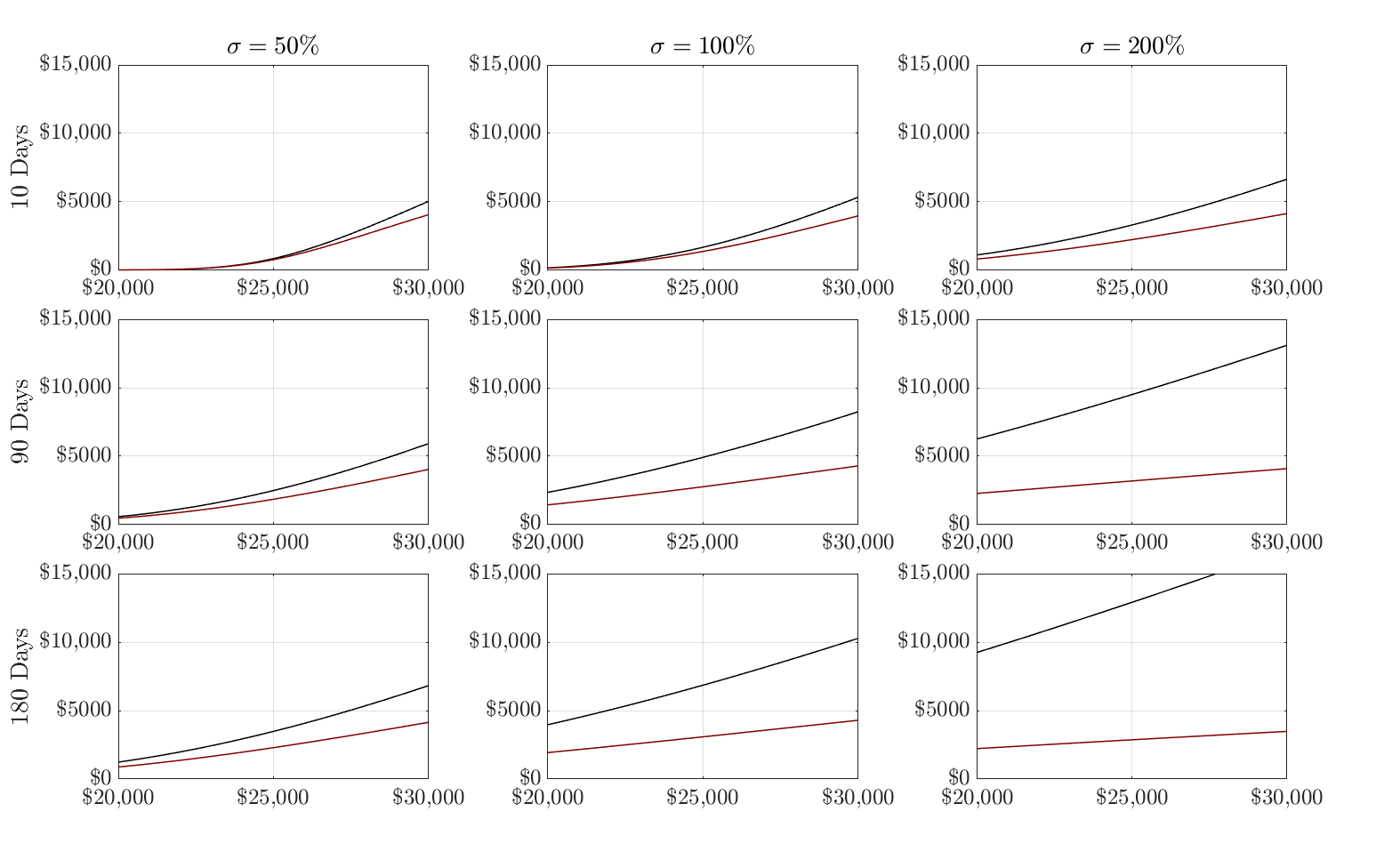}
	\label{fig:Payoff_Sigma_Tau}
	\vspace{-26pt}
\end{figure} 

Figure \ref{fig:Payoff_Sigma_Tau} illustrates inverse and quanto inverse call prices as functions of the underlying, for a given fixed strike and interest rate, and for different levels of volatility and time to maturity. Inverse option prices display the familiar pattern of increasing with either time to maturity or volatility. But quanto inverse call prices can be decreasing  with volatility (negative vega) as well as maturity (negative theta) as previously discussed. For instance, with a fixed volatility at 200\%, the fair price of a 10-day quanto inverse option with an ITM strike level at $K^{^{_{\$}}}=\$30,000$ is \$4123, but this decreases as maturity increases to 90 days (\$4080) and to 180 days (\$3490). But the OTM option with strike $K^{^{_{\$}}}=\$20,000$ is strictly increasing with maturity. The 90-day OTM quanto inverse option has positive vega whereas the ITM option has a vega which changes sign from positive to negative as the option moves deep ITM: the price increases from \$4020 ($\sigma= 50\%$) to \$4270 ($\sigma= 100\%$) and decreases afterwards to \$4080 ($\sigma= 200\%$). The  difference between ITM inverse and quanto inverse option prices is more pronounced than for OTM options, and it also increases with maturity and volatility. Especially for long-term options and/or during periods of high volatility, a quanto inverse call provides a very affordable alternative to a standard inverse call. 

\section{Hedge Ratios}\label{sec:greeks}
In the Appendix, Table \ref{tab:Greeks} summarises the pricing formulae for  inverse and quanto inverse options and their hedge ratios, under the GBM assumption. As in section \ref{sec:fxp} we assume $r^{\mathbb{B}}=0$ and set $r^{\$}=r$ for brevity, this way we have the same  interest rate for both inverse and quanto inverse options. In the Appendix, Figures \ref{fig:Greeks1Call} and \ref{fig:Greeks2Call} illustrate the delta, gamma, vega, theta, volga and vanna of inverse and quanto inverse calls as a function of strike $K^{^{_{\$}}}$ for fixed $S$, $\sigma$ and $r$ and for different time to maturity. We set the underlying volatility to $75\%$ and we compare the inverse options hedge ratios in blue with the corresponding ratios for the quanto inverse option in red, for options of different maturities (10 days, 30 days and 90 days). Figures \ref{fig:Greeks1Put} and \ref{fig:Greeks2Put}, also in the Appendix, illustrate the same  for puts.  

First compare the deltas. The inverse call (put) delta has the usual shape of a monotonically increasing (decreasing) normal distribution function. The inverse call gamma and vega are both positive, following the standard normal density and identical for call and puts. But the quanto inverse call delta is not monotonic but has a maximum when the underlying price just exceeds the strike and declines thereafter eventually becoming zero for deep ITM options. This is because the payoff is capped at $\Bar{X}$, i.e. even for large price movements the change in the payoff is limited, as has already been discussed above when commenting on the shift from convexity to concavity in the pricing function. 

Compare the impact of regular rebalancing of a simple delta-hedged position on a deep ITM direct (or standard) call versus the quanto inverse call of the same moneyness. Very frequent rebalancing of a delta-hedged position can have an adverse impact on the volatility of the underlying, especially for deep ITM calls or puts where the delata is close to $\pm 1$.\footnote{See for instance \cite{GJ2012} or  \cite{NPPW2021}}. The delta is so much greater on ITM direct (or standard) calls than it is for quanto inverse calls of the same moneyness, indeed the delta approaches zero for very very deep ITM quanto inverse calls. Therefore the unwanted volatility impact will be very much less. The non-monotonicity in the call delta is reflected in the shape of quanto inverse gamma. Comparing the term structure in the two call deltas, short-term OTM (ITM) inverse call deltas are smaller (larger) than their long-term counterparts. The quanto inverse call deltas are similar except that long-term deltas exceed the short-term for deep ITM options.   Quanto inverse call gammas have a similar shape to inverse call gamma at high (OTM) strikes but the gamma becomes negative as the strike decreases, before eventually converging to zero as the strike tends to zero. This is due to the change from a convex to concave pricing curve. 

The quanto inverse put deltas \textit{are} monotonic, as are the inverse put deltas, but the quanto inverse deltas decrease faster as a function of strike, meaning that their gamma is greater than the gamma for inverse options at strikes above the ATM strike. Like standard gamma decreases as time to expiry increases, t gradually decrease afterwards without a lower boundary. The consequences become severe for hedging, i.e. a writer would need to buy more units of the underlying, exceeding the notional amount of the option to be delta-neutral. 

By contrast with deep ITM quanto inverse calls, a deep ITM quanto inverse put delta can be very much \textit{greater} than the delta of  a direct (or standard) put of the same moneyness. Thus, the adverse volatility impact of rapid delta-hedge rebalancing referred to above \citep{GJ2012, NPPW2021} would be exacerbated. However, the delta hedge of a long position on such an option will require buying more than the notional on the underlying. This buying pressure could then cause the underlying price to rise, and so the put becomes less ITM and its delta would then decline. Finally, we note that long-term quanto inverse put deltas are generally lower than their short-term counterparts independent of the moneyness level with exceptions for short- and mid-term maturities around the ATM level. 

We have already discussed the negative theta for inverse and quanto inverse options. Figures \ref{fig:Greeks2Call} and \ref{fig:Greeks2Put} show that the only inverse options with a positive theta are in fact low strike (ITM) quanto inverse calls. Another notable feature from these figures is that the volga for longer-maturity quanto inverse puts can very large and positive, and that the vanna can take either sign. For inverse calls and puts it is negative for low-strike options and positive for high-strike options. For low-strike quanto inverse calls and puts the vanna is negative but for high-strike calls it is positive and for most high-strike puts it is negative.

\section{Conclusions}\label{sec:conc}
Developments in blockchain are driven by computer scientists who bring a fresh approach to traditional solutions and, as use of  this technology grows, traditional finance institutions are increasingly drawn towards the new and original products and trading tools now available in centralised and decentralised financial markets.  The crypto options market in particular has been expanding extremely rapidly during the last few years. Well over 90\% of global trading volumes are on inverse options, a new type of product which provides the solution for non-fiat platforms that wish to list crypto-fiat trading pairs. 

We have highlighted basis risk as main source of market incompleteness for inverse options, and described the issues arising under an indifference pricing framework.  The absence of  a
hedge portfolio leads to major difficulties so instead the pricing problem becomes that of finding an optimal trading strategy. But prices and hedge ratios then depend entirely on the trader's preferences, and whether the position is long or short. This is an interesting avenue for further research but it is not the focus of this paper.  Instead we analyse the pricing and hedging of these options under the GBM assumption. At the time of writing, all inverse options are for trading bitcoin, ether or solana against the U.S. dollar, so the GBM assumption is certainly not realistic. A similar comment applies to \textit{every}  options market, but Black-Scholes-type pricing formulae are not irrelevant. They are used to back out implied volatilities and many professional traders use the Black-Scholes hedge ratios to balance their options books for delta-gamma-vega neutrality. 

We have discussed the attractions of quanto direct options, which allow USD-denominated traders to hedge the risk of a stablecoin decoupling from its USD peg. Of course, stablecoin prices are much less variable than other crypto assets, and their de-pegging from USD is more of an operational risk than a market risk. Nevertheless, the liquidity rug pull which precipitated the Terra collapse in May 2022 shows how easy it is to attack any stablecoin that has substantial liquidity in pools of decentralised exchanges. The risks of denominating crypto trades in stablecoins has become very apparent to fiat-based traders, as well as their regulators. Quanto direct options offer a protection against stablecoin collapse which should therefore be attractive to traders on Binance and other non-fiat exchanges offering direct option products.

As well as deriving Black-Scholes hedge ratios for inverse options, we compare their pricing and hedging formulae to those of a new type of currency protected option, which we call a quanto inverse option, based on the assumption that the underlying price follows a GBM. These are not exchange-traded products at the time of writing, but we have every expectation they soon will be because USD-denominated traders can hedge the currency risk of bitcoin, ether or indeed any other crypto asset, by fixing the price of the crypto asset-USD trading pair up front. We argue that  quanto inverse options provide an attractive alternative to standard options for USD-denominated traders. Particularly when the underlying is volatile, as it is bitcoin and even more so for other cryptocurrencies and crypto assets, the profits from both long and short positions on quanto inverse puts can be much greater than they are for standard or direct puts of similar moneyness and maturity. Of course, this depends on moneyness, but also on the fix for the quanto factor.\\

\noindent We conclude by summarizing  the main points to take away from this paper:
\begin{itemize}
\item {Inverse options, which account for over 90\% of trading on crypto options, have the same payoff structure as a standard FX option, and should therefore be priced as such \citep{GK1983} even if one side of the trading pair is regarded as a security; }

\item {Nevertheless, the Deribit options market, which virtually monopolises the trading on inverse options is theoretically incomplete. Accounting for this incompleteness requires a preference-based approach, 
 under which  the option price would depend on the trader's utility, risk tolerance and perhaps even his initial wealth;}

\item {We explain how quanto direct options can offer all traders protection against decoupling of a stablecoin (such as tether) price from USD.}

\item {Another use case for quanto crypto options is to provide exposure to another crypto without traders needing to change their base currency. For instance, a BTC-based trader can profit from a change in the XRP price by fixing the BTC/USDT rate in a direct quanto on XRP/USDT, whereby the direct option payoff in USDT is paid out in BTC. This way, there is also no need for fiat onboarding.}

\item {Quanto inverse options pose a new type of exotic option which allow USD-denominated traders to gain exposure to the expanding crypto market without taking any crypto on the balance sheet. For instance, a trader could fix a BTC/USD quanto factor, so that any options that are settled in BTC have all profit and loss automatically converted to USD.  }

\item {The concave (call) and convex (put) payoff structure of quanto inverse options possess features that are attractive to both buyers and sellers. The uncapped USD put put payoff is a perfect insurance against crypto price crashes. The capped USD payoff for a quanto inverse call results in prices that are lower than standard (or direct) calls of the same moneyness and maturity, depending on the conversion factor.}

\item All Deribit options are accounted in BTC, ETH or SOL, and only these currencies can be used for deposits and withdrawals. But it takes time for  USD-based traders to convert between USD and these crypto, perhaps using a US-based exchange like Coinbase or Kraken, especially during periods of high volatility. This friction induces a small, but non-zero liquidity premium on the prices of Deribit inverse options. The quanto inverse option avoids such a liquidity premium.

\end{itemize}

\bibliographystyle{chicago}
\bibliography{Paper}

\begin{appendix}
\counterwithin{figure}{section}
\counterwithin{table}{section}
\section*{Appendix A: Inverse and Quanto Inverse Option Prices and Greeks}

\begin{table}[h!]
			\centering
			  \renewcommand\thetable{A.1}
			\caption{Inverse and Quanto Inverse Option Prices and Greeks}
			\caption*{\footnotesize  We assume the tradable underlying price $S$ follows a GBM with volatility $\sigma$, where the drift depends on the discount rate $r$.  The inverse or quanto inverse call or put have strike $K$ and for the sake if clarity we omit the pre-defined $\Bar{X}$ for the quanto inverse case. Note that this need to be multiplied to the individual Greek. The residual time to maturity $\tau$ hold for all types and we use the notation $\omega  = \pm 1$ according as the option is a call or a put. Suppressing all time subscripts for simplicity we set  $d_1 = \ln \left(\frac{S}{K}\right)\left[\sigma \sqrt{\tau}\right]^{-1}+ ( r + \frac{1}{2}\sigma )\sqrt{\tau}$, $d_2 = d_1 - \sigma \sqrt{\tau}$ as well as $d_3 = d_2 - \sigma \sqrt{\tau} $.}
			\begin{tabular}{llllcc}
				\toprule
				Name  &    &   & & Inverse   & Quanto inverse      \\
				\midrule
				\midrule
				Price 
				& $f$ &  & 
				&  $\omega [ S \Phi(\omega d_1) - e^{-r\tau}K \Phi (\omega d_2)  ] $  
				& $\omega  e^{-r\tau}[\Phi(\omega d_2) - e^{(  \sigma^2-r)\tau }S^{-1}K \Phi(\omega d_3)] $
				\\
					[1.5em]

				Delta 
				& $\delta$ &       
				& $ \frac{\partial f}{\partial S}$ 
				& $\omega  \Phi(\omega d_1)$  
				& $ \omega e^{  (\sigma^2-2r) \tau } \frac{K}{S^2} \Phi(\omega d_3) $  
				  \\
				[1.5em]
				
				Gamma  
				& $\gamma$&       
				& $\frac{\partial^2 f}{\partial S^2}$ 
				& $ \frac{{ \phi}(d_1)}{S \sigma \sqrt{\tau}}$     
				&  $ e^{ (\sigma^2 -2r)  \tau} \frac{K}{S^3} \left[   \frac{ \phi(d_3)}{\sigma \sqrt{\tau}} -\omega2\Phi(\omega d_3) \right] $     	
				 \\
				[1.5em]
				
				Vega  
				& $\nu$ &       
				& $\frac{\partial f}{\partial \sigma}$ 
				&  $S \phi(d_1) \sqrt{\tau}$    
				&  $ e^{-r\tau} \left[ \phi(d_2) \sqrt{\tau} - \omega  e^{(\sigma^2-r) \tau} \sigma \tau \frac{2K }{S} \Phi(\omega d_3) \right] $    
				  \\
				[3em]
				
				Volga 
				& $\nu^\text{\tiny{$o$}}$  &       
				& $\frac{\partial^2 f}{\partial \sigma^2} $  
				& $   \sqrt{\tau} \phi(d_1) S \frac{d_1 d_2}{\sigma}  $    
				& $\!\begin{aligned}[t] & e^{-r\tau} \sqrt{\tau}\phi(d_2)  \frac{d_2 d_1}{\sigma} -\omega 2\frac{K}{S}\tau 	e^{\left(\sigma^2 -2r \right)\tau} \times \\
				&\left[ \Phi(\omega d_3)(1+2\sigma^2 \tau)-\sigma \phi(d_3)\left(\frac{-d_1}{\sigma}-\sqrt{\tau} \right)\right]
				\end{aligned}$
				     \\
				[5em]
				
				Vanna &  $\nu^\text{\tiny{$a$}}$ &       
				& $\frac{\partial^2 f}{\partial S\partial \sigma}$  
				& $ - \phi(d_1) \frac{d_2}{\sigma}$ 
				& $\!\begin{aligned}[t] & \qquad \omega e^{(\sigma^2-2r) \tau} \times \\
				& \frac{K}{S^2} \left[ 2 \tau \sigma\Phi(\omega d_3)  + \omega \phi(d_3) \left(\frac{-d_1}{\sigma} - \sqrt{\tau}\right) \right] 	
				 \end{aligned} $    
				        \\
				[5em]
				
				Theta &  $\vartheta$ &       
				&    $-\frac{\partial f}{\partial \tau}$ 
				&   $\!\begin{aligned}[t] &\omega[-re^{-r\tau} K\Phi(\omega d_2) ] \\
				&- \frac{\sigma S \phi(d_1) }{2\sqrt{\tau}}\end{aligned}$
				& $\!\begin{aligned}[t] & -e^{-r\tau} \left(\frac{\phi(d_2)\sigma}{2\sqrt{\tau} } - \omega r \Phi(\omega d_2)\right)\\
				&  + \omega e^{(\sigma^2 -2r)\tau}\frac{K}{S}(\sigma^2 -2r) \Phi(\omega d_3) \end{aligned}$         \\

			\bottomrule
			\end{tabular}%
			\label{tab:Greeks}%
	\end{table}%

\clearpage

\section*{Appendix B: Derivation of the Greeks}

\begin{align*}
  d_2 &= \frac{\ln \left(\frac{S^{\$}_{_T}}{K}\right)+\left( r- \frac{1}{2}\sigma^2\right)\tau}{\sigma \sqrt{\tau}} \qquad   
   d_3 = \frac{\ln \left(\frac{S^{\$}_{_T}}{K}\right)+\left( r- \frac{3}{2}\sigma^2\right)\tau}{\sigma \sqrt{\tau}} = d_2-\sigma\sqrt{\tau}
\end{align*}

\begin{align*}
    \phi(d_2) & = \phi(-d_2)=  \frac{1}{\sqrt{2 \pi}} e^{-\frac{1}{2} d_2^2} \\
    \phi(d_3) &= \phi\left(d_2-\sigma\sqrt{\tau}\right) =\frac{1}{\sqrt{2 \pi}} e^{-\frac{1}{2} \left(d_2-\sigma\sqrt{\tau}\right)^2} =\frac{1}{\sqrt{2 \pi}} e^{-\frac{1}{2} \left(d^2_2- 2d_2\sigma\sqrt{\tau} +\sigma^2 \tau \right)}\\   
    &=\frac{1}{\sqrt{2 \pi}} e^{-\frac{1}{2} d^2_2} e^{d_2\sigma\sqrt{\tau}} e^{-\frac{1}{2}\sigma^2 \tau } =\phi(d_2) e^{\ln \left(\frac{S^{\$}_{_T}}{K}\right)+\left( r- \frac{1}{2}\sigma^2\right)\tau} e^{-\frac{1}{2}\sigma^2 \tau }\\ 
    &=\phi(d_2) \frac{S^{\$}_{_T}}{K} e^{\left( r- \sigma^2\right)\tau} 
\end{align*}

\begin{align*}
\frac{\partial d_2}{\partial F} = \frac{\partial d_3}{\partial F} = \frac{1}{F \sigma \sqrt{\tau}}
\end{align*}

\begin{align*}
\frac{\partial d_2}{\partial \sigma} = \frac{-\ln \left(\frac{S^{\$}_{_T}}{K}\right) -r\tau - \frac{1}{2}\sigma^2\tau}{\sigma^2 \sqrt{\tau}} \qquad   
  \frac{\partial d_3}{\partial \sigma} = \frac{-\ln \left(\frac{S^{\$}_{_T}}{K}\right) -r\tau - \frac{3}{2}\sigma^2\tau}{\sigma^2 \sqrt{\tau}} = \frac{\partial d_2}{\partial \sigma}-\sqrt{\tau}
\end{align*}

\begin{align*}
\frac{\partial d_2}{\partial \tau} =  \frac{-0.5\ln \left(\frac{S^{\$}_{_T}}{K}\right) +0.5r\tau - \frac{1}{4}\sigma^2\tau}{\sigma \sqrt{\tau^3}} \qquad   
  \frac{\partial d_3}{\partial \tau} =  \frac{-0.5\ln \left(\frac{S^{\$}_{_T}}{K}\right) +0.5r\tau - \frac{3}{4}\sigma^2\tau}{\sigma \sqrt{\tau^3}} = \frac{\partial d_2}{\partial \tau}-\frac{\sigma}{2\sqrt{\tau}}
\end{align*}

\begin{align*}
{\delta} := \frac{\partial f}{\partial F}&=\frac{\partial}{\partial F} e^{-r\tau} \left[ \Phi (d_2) - e^{\left(\sigma^2-r\right) \tau } F_t ^{-1} K \Phi(d_3)\right]\\ 
&= e^{-r\tau} \left[ \frac{\partial \Phi(d_2)}{\partial d_2} \frac{\partial d_2}{\partial F} - e^{(\sigma^2-r)\tau} \left[-\frac{ K}{F^2} \Phi(d_3) + \frac{K}{F}  \frac{\partial \Phi(d_3)}{\partial d_3} \frac{\partial d_3}{\partial F}\right] \right]\\
&=  e^{-r\tau} \left[ \frac{\phi(d_2) }{F \sigma \sqrt{\tau}} - e^{(\sigma^2-r) \tau} \left[-\frac{K}{F^2} \Phi(d_3) + \frac{K}{F^2 \sigma \sqrt{\tau}}  \phi(d_3) \right] \right]\\
&=  e^{-r\tau} \left[\frac{\phi(d_2) }{F \sigma \sqrt{\tau}} + e^{(\sigma^2-r) \tau}  \frac{K}{F^2} \Phi(d_3) - e^{(\sigma^2-r)\tau}  \frac{K}{F^2 \sigma \sqrt{\tau}}  \phi(d_2-\sigma\sqrt{\tau}) \right]\\
&=  e^{-r\tau} \left[\frac{\phi(d_2) }{F \sigma \sqrt{\tau}} + e^{(\sigma^2-r) \tau}  \frac{K}{F^2} \Phi(d_3) - e^{(\sigma^2-r)\tau}  \frac{K}{F^2 \sigma \sqrt{\tau}} \phi(d_2) \frac{F}{K} e^{-(\sigma^2-r)\tau}\right]\\
&=e^{(\sigma^2 -2r)\tau} F^{-2}K \Phi(d_3)
\end{align*}

\begin{align*}
{\gamma} := \frac{\partial \delta}{\partial F}&=\frac{\partial}{\partial F} \left[ e^{(\sigma^2 -2r)\tau} F^{-2}K \Phi(d_3) \right]
=e^{(\sigma^2-2r) \tau} F^{-3}K \left[ \frac{\phi(d_3)}{\sigma \sqrt{\tau}} -2\Phi(d_3) \right]
\end{align*}

\begin{align*}
\nu := \frac{\partial f}{\partial \sigma}&=\frac{\partial}{\partial \sigma}  e^{-r\tau} \left[ \Phi (d_2) - e^{\left(\sigma^2-r\right) \tau } F_t ^{-1} K \Phi(d_3)\right]\\ 
&= e^{-r\tau} \left[ \frac{\partial \Phi(d_2)}{\partial d_2} \frac{\partial d_2}{\partial \sigma} - \frac{K}{F} \left[ \frac{\partial e^{(\sigma^2-r) \tau}}{\partial \sigma} \Phi(d_3) +e^{(\sigma^2-r) \tau}  \frac{\partial \Phi(d_3)}{\partial d_3} \frac{\partial d_3}{\partial \sigma}\right] \right]\\
& =  e^{-r\tau} \left[\phi(d_2) \frac{\partial d_2}{\partial \sigma} - e^{(\sigma^2 -r)\tau} \frac{K}{F} \left[ 2 \tau \sigma  \Phi(d_3) + \phi(d_2 - \sigma \sqrt{\tau}) \left(\frac{\partial d_2}{\partial \sigma} - \sqrt{\tau}\right)  \right] \right] \\
& =  e^{-r\tau} \left[\phi(d_2) \frac{\partial d_2}{\partial \sigma} - e^{(\sigma^2 -r)\tau} \frac{K}{F} \left[ 2 \tau \sigma  \Phi(d_3) + \phi(d_2) e^{d_2 \sigma \sqrt{\tau}} e^{-\frac{\sigma^2 \tau}{2} } \left(\frac{\partial d_2}{\partial \sigma} - \sqrt{\tau}\right)  \right] \right] \\
& =  e^{-r\tau} \left[\phi(d_2) \frac{\partial d_2}{\partial \sigma} - e^{(\sigma^2 -r)\tau} \frac{K}{F} \left[ 2 \tau \sigma  \Phi(d_3) + \phi(d_2) \frac{F}{K} e^{-(\sigma^2-r)\tau} \left(\frac{\partial d_2}{\partial \sigma} - \sqrt{\tau}\right)  \right] \right] \\
& =  e^{-r\tau} \left[ \phi(d_2)\sqrt{\tau} - 2 e^{(\sigma^2-r) \tau} F^{-1}K \tau \sigma \Phi(d_3) \right]
\end{align*}

\begin{align*}
\vartheta := \frac{\partial f}{\partial \tau}&=\frac{\partial}{\partial \tau}  e^{-r\tau}\Phi (d_2) - e^{\left(\sigma^2-2r\right) \tau } F_t ^{-1} K \Phi(d_3)\\ 
&= \left[  \frac{\partial e^{-r\tau}}{\partial \tau} \Phi(d_2) + e^{-r\tau} \frac{\partial \Phi(d_2)}{\partial d_2} \frac{\partial d_2}{\partial \tau} \right] - \frac{K}{F} \left[ \frac{\partial e^{(\sigma^2-2r) \tau}}{\partial \tau} \Phi(d_3) +e^{(\sigma^2-r) \tau}  \frac{\partial \Phi(d_3)}{\partial d_3} \frac{\partial d_3}{\partial \tau}\right] \\
& = \left[-re^{-r\tau}\Phi (d_2) + e^{-r\tau} \phi(d_2)\frac{\partial d_2}{\partial \tau} \right] - \frac{K}{F} \left[\left(\sigma^2 -2r\right) \Phi (d_3) e^{(\sigma^2-2r)\tau} + e^{(\sigma^2-2r) \tau} \phi(d_3)\frac{\partial d_3}{\partial \tau}   \right] \\
&= \left[-re^{-r\tau}\Phi (d_2) + e^{-r\tau} \phi(d_2)\frac{\partial d_2}{\partial \tau} \right] - \\
& \qquad \qquad \qquad \frac{K}{F} \left[\left(\sigma^2 -2r\right) \Phi (d_3)e^{(\sigma^2-2r)\tau} + e^{(\sigma^2-2r) \tau} \phi(d_2-\sqrt{\tau})\left( \frac{\partial d_2}{\partial \tau} - \frac{\sigma}{2 \sqrt{\tau}} \right)   \right]\\
&= \left[-re^{-r\tau}\Phi (d_2) + e^{-r\tau} \phi(d_2)\frac{\partial d_2}{\partial \tau} \right] - \\
& \qquad \qquad \qquad \frac{K}{F} \left[\left(\sigma^2 -2r\right) \Phi (d_3)e^{(\sigma^2-2r)\tau} + e^{(\sigma^2-2r) \tau} \phi(d_2) \frac{S^{\$}_{_T}}{K} e^{\left( r- \sigma^2\right)\tau} \left( \frac{\partial d_2}{\partial \tau} - \frac{\sigma}{2 \sqrt{\tau}} \right)   \right]\\
&=\left[-re^{-r\tau}\Phi (d_2) + e^{-r\tau} \phi(d_2)\frac{\partial d_2}{\partial \tau} \right] -  \left(\sigma^2 -2r\right)\Phi (d_3) \frac{K}{F} e^{(\sigma^2-2r)\tau}-  e^{-r\tau} \phi(d_2) \left( \frac{\partial d_2}{\partial \tau} - \frac{\sigma}{2 \sqrt{\tau}} \right) \\
&=-re^{-r\tau}\Phi (d_2) + e^{-r\tau} \phi(d_2)\frac{\partial d_2}{\partial \tau}  -  \left(\sigma^2 -2r\right)\Phi (d_3) \frac{K}{F}e^{(\sigma^2-2r)\tau} -  e^{-r\tau} \phi(d_2) \frac{\partial d_2}{\partial \tau}  +   e^{-r\tau} \phi(d_2)\frac{\sigma}{2 \sqrt{\tau}}\\
&=-re^{-r\tau}\Phi (d_2) - e^{(\sigma^2-2r)\tau} \left(\sigma^2 -2r\right)\Phi (d_3) \frac{K}{F} +   e^{-r\tau} \phi(d_2)\frac{\sigma}{2 \sqrt{\tau}}\\
\end{align*}
\clearpage

\section*{Appendix C: Inverse and Quanto Inverse Greeks}

\begin{figure}[h!]
	\centering
	\small
	\renewcommand{\thefigure}{C.1}
	\caption{Inverse and Quanto Inverse Call Greeks I}
	\vspace{-0.3cm}
	\caption*{\footnotesize  Inverse and quanto inverse Greeks for calls as a function of the strike level $K^{^{_{\$}}}$ with fixed underlying $S^{^{_{\$}}}_{_t}$ and $\Bar{X}=S^{^{_{\$}}}_{_t}=\$25,000$, volatility $\sigma=75\%$ and different times to maturity: from 10 days (bold), 30 days (dashed) and 90 days (dotted).  The left, blue column represents the inverse Greeks, while the right, red column shows the quanto inverse Greeks.}
	\includegraphics[width=\textwidth]{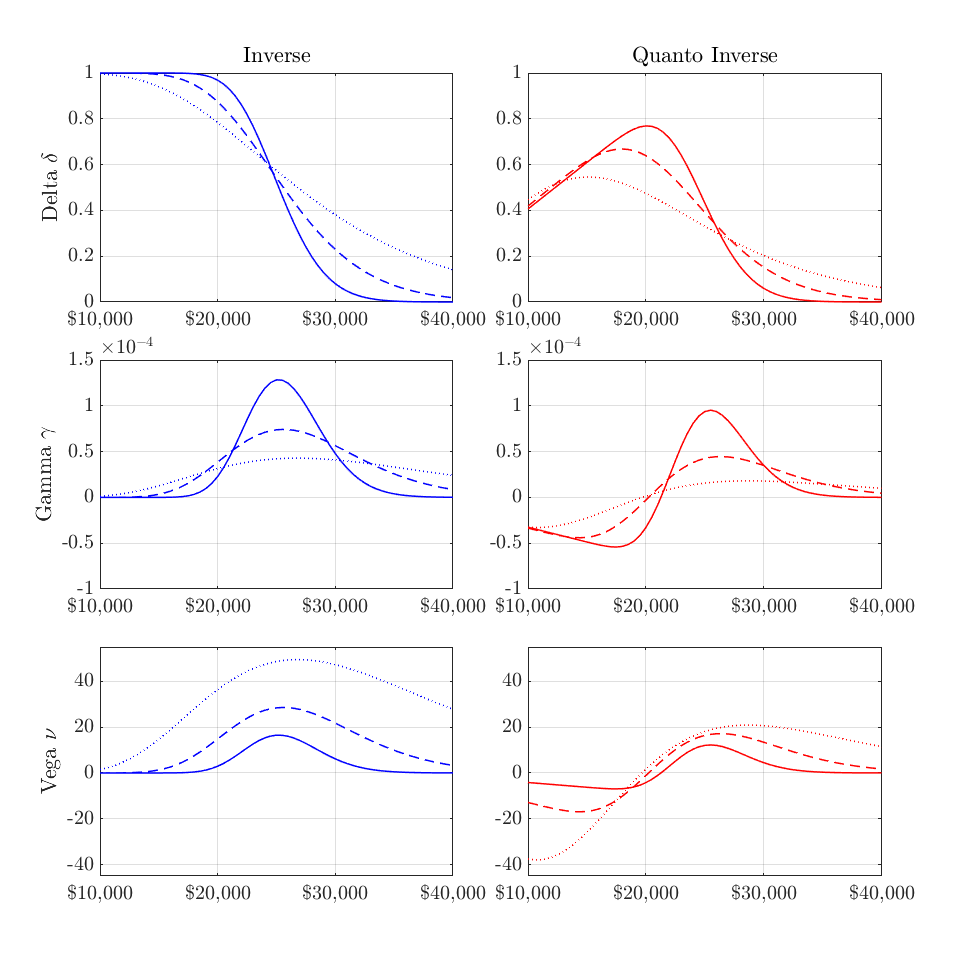}
	\label{fig:Greeks1Call}
\end{figure} 

\clearpage

\begin{figure}[h!]
	\centering
	\small
	\renewcommand{\thefigure}{C.2}
	\caption{Inverse and Quanto Inverse Call Greeks II}
	\vspace{-0.3cm}
	\caption*{\footnotesize  Inverse and quanto inverse Greeks for calls as a function of the strike $K$ with fixed underlying and predetermined conversion factor $S^{^{_{\$}}}_{_t}=\Bar{X}=\$25,000$, volatility $\sigma=75\%$ and different times to maturity: from 10 days (bold), 30 days (dashed) and 90 days (dotted).  The left, blue column represents the inverse Greeks, while the right, red column displays the quanto inverse Greeks.}
	\includegraphics[width=\textwidth]{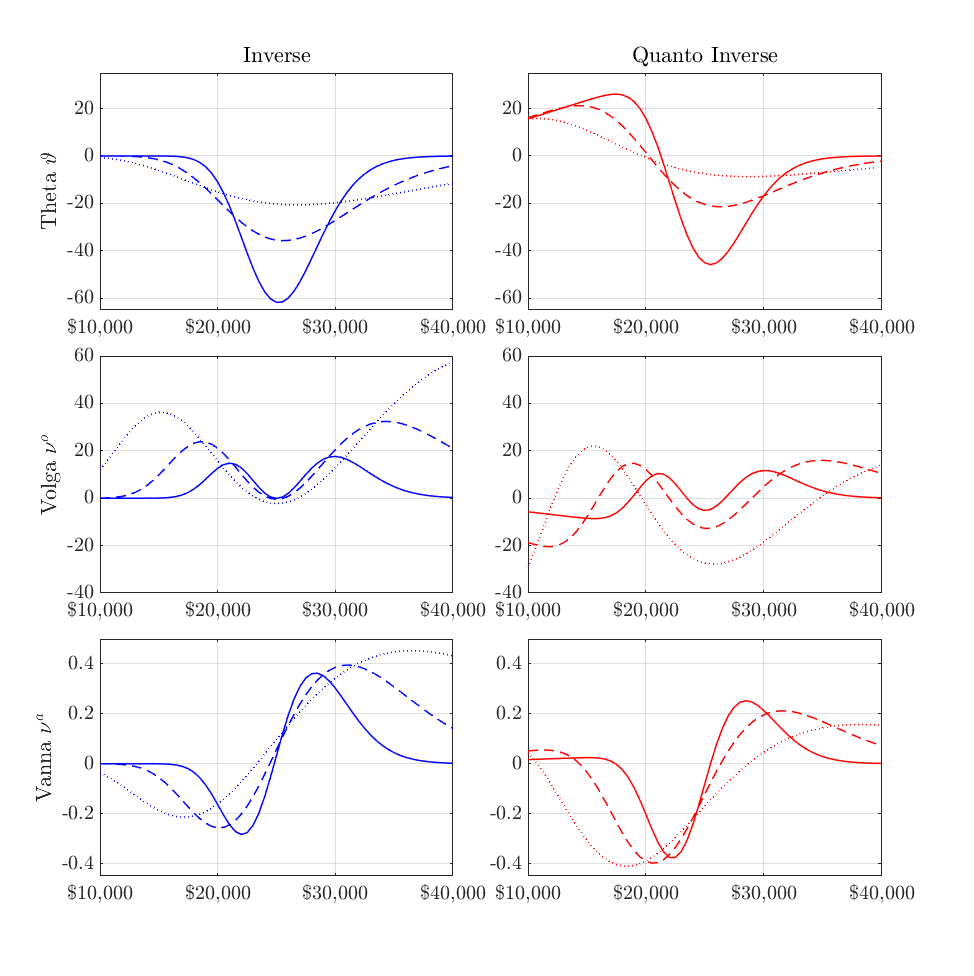}
	\label{fig:Greeks2Call}
\end{figure} 

\clearpage

\begin{figure}[h!]
	\centering
	\small
	\renewcommand{\thefigure}{C.3}
	\caption{Inverse and Quanto Inverse Put Greeks I }
	\vspace{-0.3cm}
	\caption*{\footnotesize  Inverse and quanto inverse Greeks for puts as a function of the strike $K$ with fixed underlying and predetermined conversion factor $S^{^{_{\$}}}_{_t}=\Bar{X}=\$25,000$, volatility $\sigma=75\%$ and different times to maturity: from 10 days (bold), 30 days (dashed) and 90 days (dotted).  The left, blue column represents the inverse Greeks, while the right, red column displays the quanto inverse Greeks.}
	\includegraphics[width=\textwidth]{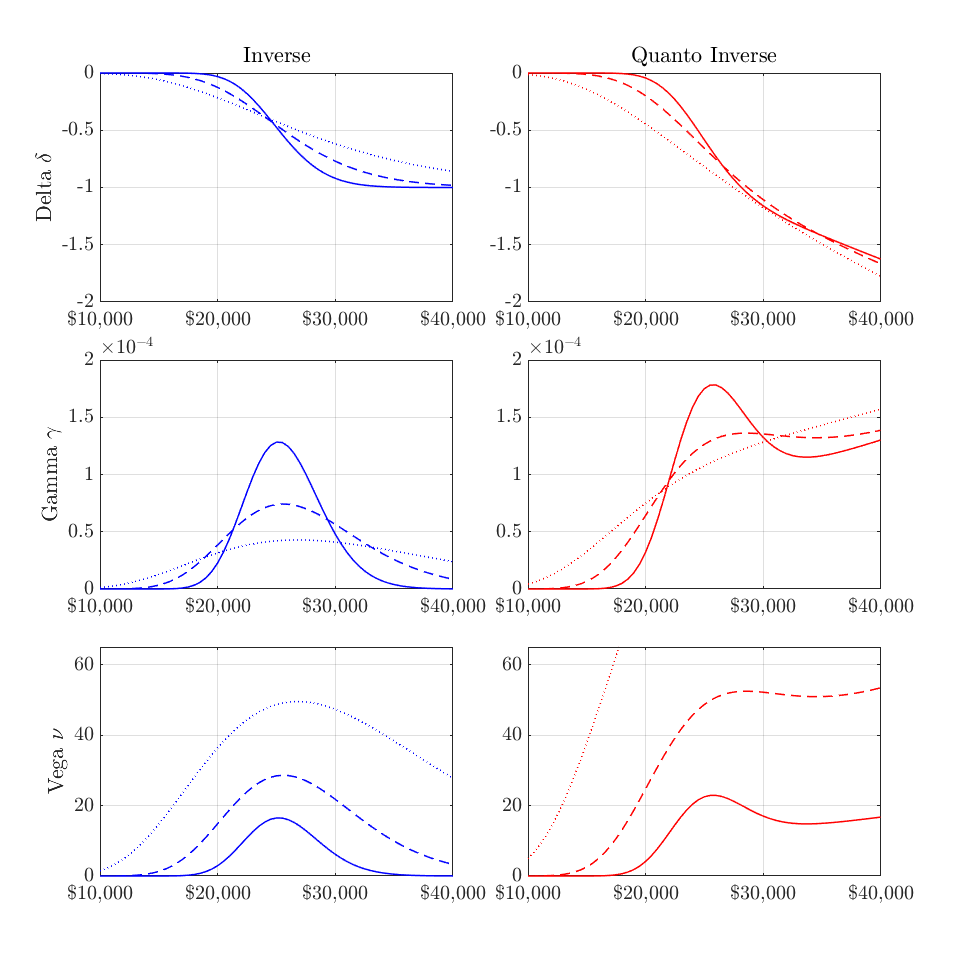}
	\label{fig:Greeks1Put}
\end{figure} 

\clearpage

\begin{figure}[h!]
	\centering
	\small
	\renewcommand{\thefigure}{C.4}
	\caption{Inverse and Quanto Inverse Put Greeks II }
	\vspace{-0.3cm}
	\caption*{\footnotesize   Inverse and quanto inverse Greeks for puts as a function of the strike $K$ with fixed underlying and predetermined conversion factor $S^{^{_{\$}}}_{_t}=\Bar{X}=\$25,000$, volatility $\sigma=75\%$ and different times to maturity: from 10 days (bold), 30 days (dashed) and 90 days (dotted).  The left, blue column represents the inverse Greeks, while the right, red column displays the quanto inverse Greeks.}
	\includegraphics[width=\textwidth]{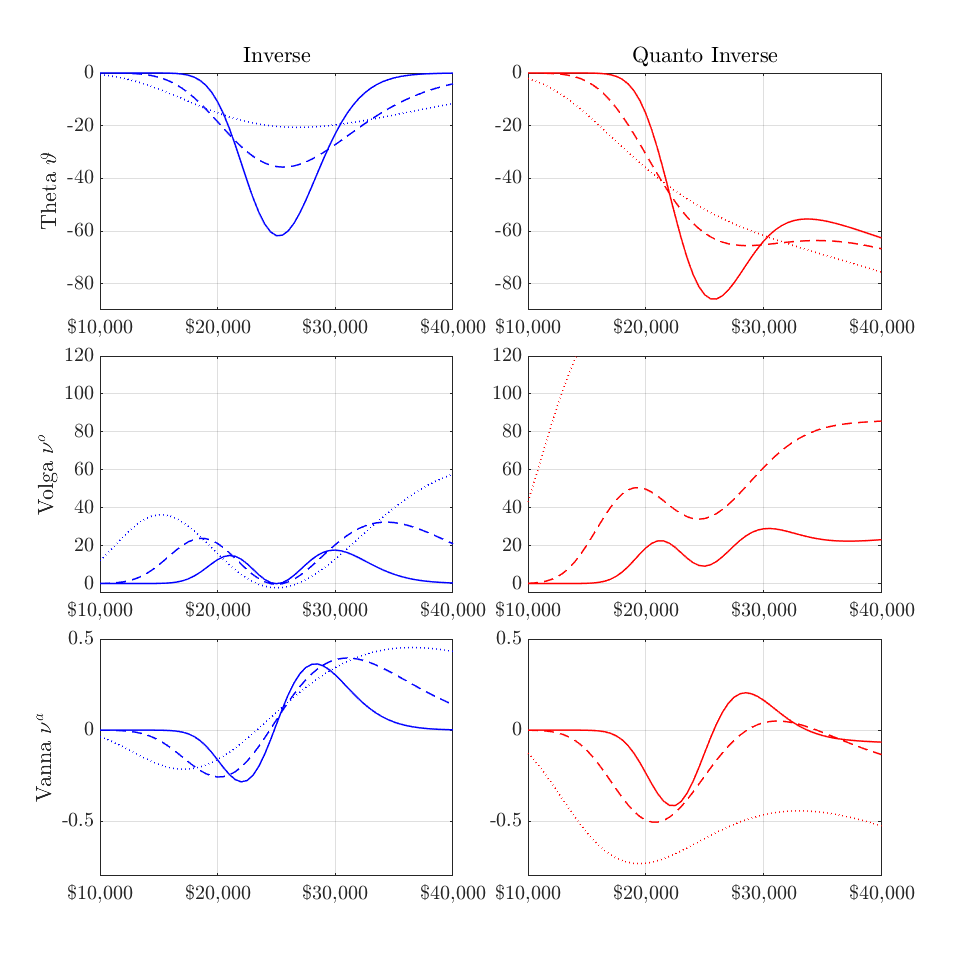}
	\label{fig:Greeks2Put}
\end{figure}

\end{appendix}

\end{document}